\documentclass[floatfix,nofootinbib,12pt,prx]{revtex4}
\usepackage[utf8]{inputenc}
\usepackage{graphicx,amsfonts,amssymb,amsmath,hyperref,hypcap,enumerate}
\usepackage{slashed}
\usepackage{amsmath,amssymb,bm,graphicx,dsfont}
\usepackage{amsmath}
\usepackage{slashed}
\usepackage{dsfont}
\usepackage{amstext}
\usepackage{amssymb}
\usepackage{amsbsy}
\usepackage{amsthm}
\usepackage{epsfig}
\usepackage{graphicx}
\usepackage{braket}

\usepackage{bbm}
\usepackage{hyperref}
\usepackage{color}
\usepackage{xcolor}
\usepackage{verbatim}
\usepackage{slashed}
\renewcommand{\]}{\right]}
\renewcommand{\v}[1]{\mathbf{#1}} 

\newcommand{\be}{\begin{equation}}
\newcommand{\ba}{\begin{align}}
\newcommand{\ee}{\end{equation}}
\newcommand{\bea}{\begin{eqnarray}}
\newcommand{\eea}{\end{eqnarray}}
\newcommand{\beq}{\begin{equation}}
\newcommand{\eeq}{\end{equation}}
\newcommand{\beqn}{\begin{eqnarray}}
\newcommand{\eeqn}{\end{eqnarray}}

\usepackage{graphicx}
\usepackage{dcolumn}
\usepackage{bm}

\definecolor{darkgreen}{rgb}{0.0, 0.5, 0.0}

\begin{document}

\title{Evolution between  quantum Hall and conducting phases: simple models and some results }
\author{Zhihuan Dong}
\affiliation{Department of Physics, Massachusetts Institute of
Technology, Cambridge, MA 02139, USA}
 \author{T. Senthil}
\affiliation{Department of Physics, Massachusetts Institute of
Technology, Cambridge, MA 02139, USA}
\date{\today}

\begin{abstract}
Quantum many particle systems in which the kinetic energy, strong correlations, and band topology are all important  pose an interesting and topical challenge. Here we introduce and study particularly simple models where all of these elements are present. We consider interacting quantum particles in two dimensions in a strong magnetic field such that the  Hilbert space is restricted to the Lowest Landau Level (LLL) . This is the familiar quantum Hall regime with rich physics determined by the particle filling and statistics. A periodic potential with a unit cell enclosing one flux quantum broadens the LLL into a Chern band with a finite bandwidth. The states obtained in the quantum Hall regime evolve into conducting states in the limit of large bandwidth. We study this evolution in detail for the specific case of bosons at filling factor $\nu = 1$. In the quantum Hall regime the ground state at this filling is a gapped quantum hall state (the ``bosonic Pfaffian'') which may be viewed as descending from a (bosonic) composite fermi liquid. At large bandwidth the ground state is a bosonic superfluid. We show how both phases and their evolution can be described within  a single theoretical framework based on a LLL composite fermion construction. Building on our previous work on the bosonic composite fermi liquid, we show that the evolution into the superfluid can be usefully described by a non-commutative quantum field theory in a periodic potential. 
\end{abstract}

\maketitle

\tableofcontents

\section{Introduction}
A contemporary challenge in quantum condensed matter physics is to understand many body systems where inter-particle interactions, the kinetic energy, and band topology all play a crucial role.  An important context where all three of these elements are present\cite{po2018origin,zhang2019nearly,song2018all,po2018faithful,ahn2019failure,zhang2019twisted,bultinck2020mechanism} are  moir\'e graphene structures that have been studied intensely in the last few years (see, eg, Refs. \onlinecite{balents2020superconductivity,andrei2021marvels,
cao2018correlated,cao2018unconventional,yankowitz2019tuning,
sharpe2019emergent,chen2020tunable,serlin2020intrinsic,spanton2018observation}). 
There is an active experimental effort (see, eg, Ref. \onlinecite{kang2020topological})  to identify and study other correlated materials where the bands are topological and will typically have a non-zero bandwidth which will compete with the interactions. 

There is a long history of theoretical study of situations where only two of these three ingredients are present. 
In the absence of band topology, the  competition between the kinetic energy and the inter-particle interaction in a crystalline solid is  often discussed in the framework of interacting  lattice models such as the celebrated Hubbard model.  However if the bands are topological, the passage to an interacting lattice model is complicated by the absence of well-localized Wannier functions that manifest the microscopic symmetries of the system, and a different framework is needed. A different example is the fractional quantum Hall effect and related phenomena which happen when a single Landau level is partially filled. Such a Landau level may be viewed as a particularly simple topological band, namely a Chern band with a constant Berry curvature. Quantum Hall phenomena are not usually understood through lattice Hubbard-like models but through other approaches, eg, through wavefunctions or effective field theories. Apart from the special band topology, a Landau level has the further  special feature of  the absence of energy dispersion. The kinetic energy is quenched in a Landau level and thus the physics is determined solely by interactions (and implicitly the special topology of the single particle states spanning the Landau level). Finally there is a well developed literature dealing with the physics of weakly correlated topological materials where the inter-particle interaction does not play a major role. 

\begin{figure}
    \centering
    \includegraphics[width=0.9\linewidth]{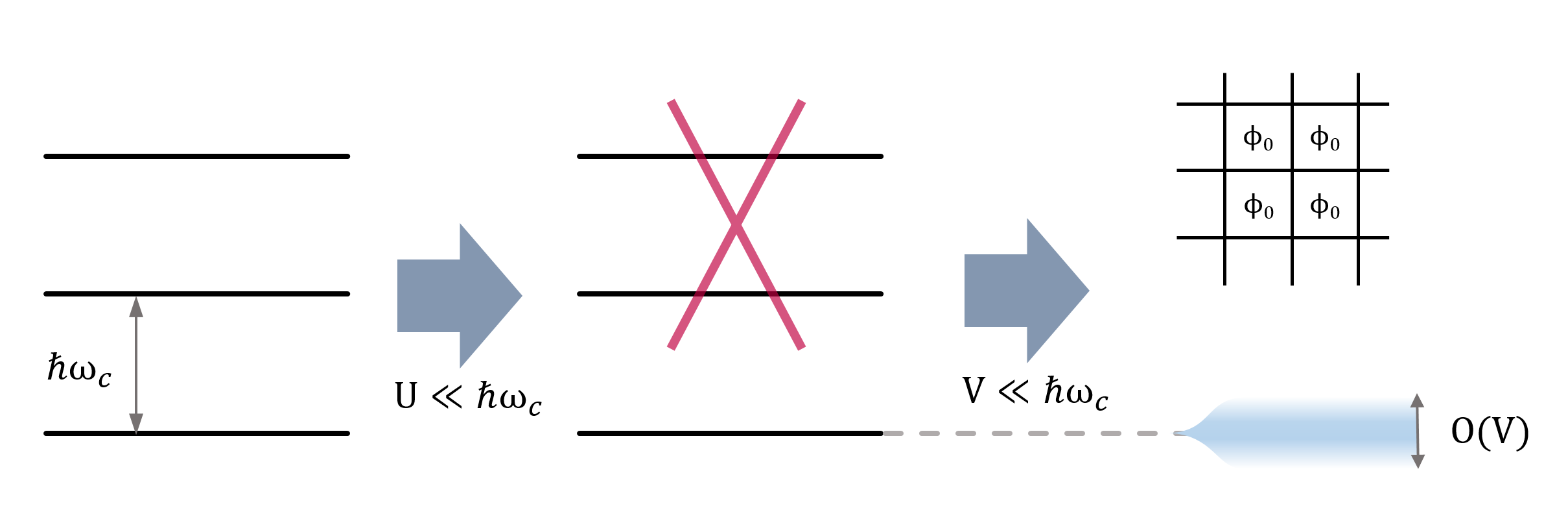}
    \caption{Projection to LLL and the effect of a periodic potential. $\omega_c$ is the cyclotron frequency. $U$ denotes the inter-particle interaction, and $V$ the strength of a periodic potential whose unit cell encloses exactly one flux quantum $\phi_0$.}
    \label{LLLprojection}
\end{figure}

In this paper we focus on a class of simple models where all three of the ingredients mentioned above are present. Consider particles (either bosons or fermions) at a mean density $\underline{\rho}$ in a strong magnetic field $B$ in spatial dimensions $d = 2$. We take the filling factor $\nu = \frac{2\pi \underline{\rho}}{B}$ to be less than $1$. We will take the limit that the Landau level spacing is much bigger than all other energy scales in the problem. For $\nu < 1$, we may then restrict the Hilbert space to be that of the lowest Landau level.  After projection  to the lowest Landau level, the Hamiltonian will be expressed in terms of a density operator $\rho^L_{\v q}$ that satisfies the well-known Girvin-MacDonald-Platzman (GMP) algebra
\be
\label{gmp}
[\rho^L_{\v q}, \rho^L_{\v p}] = 2i \sin{\frac{\v q \times \v p}{2}}\rho^L_{\v q + \v p}
\ee
We have chosen units of length such that the magnetic length $l_B^2 = \frac{1}{B} = 1$. 
The models we study have the Hamiltonian
\be
\label{Hamiltonian}
    H = H_1+H_2 = \int \frac{d^2\v q}{(2\pi)^2}~\tilde{V}(\v q)\rho^L_{-\v q} + \tilde{U}(\v q)\rho^L_{\v q}\rho^L_{-\v q}
\ee
 The  first term represents an one-body periodic potential:  we will consider the situation where  the corresponding  unit cell encloses exactly one flux quantum. A concrete example is a potential whose unit cell is a square of size $a \times a$. The flux per unit cell is then $B a^2 = a^2$ which is fixed to be $2\pi$. In momentum space correspondingly there will be reciprocal lattice vectors $\v G = \frac{2\pi}{a} (m_x,m_y)$ where $m_{x,y}$ are integers and $\tilde{V}(\v q)$ will be delta functions centered at the various $\v G$. The second term is a two-body repulsive interaction.  If in the full Hilbert space ({\em i.e} one involving all the Landau levels), there is a one-body potential $V(\v q)$ and an interaction $U(\v q)$,  then the corresponding projected potentials are
\begin{align}
    \tilde{V}(\v q)&=V(\v q)e^{-q^2/4}=\sum_{\v G}{V_{\v G}\delta^2(\v q-\v G)}e^{-q^2/4}\\
    \tilde{U}(\v q)&=U(\v q)e^{-q^2/2}
\end{align}
We parametrize the strength of the $\tilde{U}(\v q)$ by $U_0$ and that of $\tilde{V}$ by $V_0$. What is the phase diagram as a function of $\frac{V_0}{U_0}$ at some fixed $\nu$? 

For $V_0 = 0$, Eqn.~\ref{Hamiltonian} defines the familiar physics of the quantum hall regime in the lowest Landau level. For example, if the particles are fermions and $\nu = \frac{1}{3}$ we get the Laughlin $1/3$ quantum Hall state. Clearly such a fractional quantum hall state is stable to turning on a small $V_0$. 

For $U_0 = 0$, we get a free particle model. The one-body potential splits the degeneracy of the Landau level and leads to the formation of a Chern band with Chern number $1$. This is most simply understood by recognizing that though the $V$ term breaks the continuous magnetic translation symmetry of the continuum Landau level, it preserves the symmetry of discrete magnetic translations by $a$ along the $x$ or $y$ axis. These discrete translations commute with each other (as the flux per unit cell is $2\pi$). Thus they may be simultaneously diagonalized, and their common eigenstates form a complete basis for the single particle Hilbert space of the lowest Landau  level. They are  also eigenstates of the projected periodic potential. These states have a non-zero crystal momentum $\v k$ and will have an energy $\epsilon(\v k)$ that is proportional  to $V$. Thus the periodic potential  gives the Landau level a finite bandwidth. However since it merely changes the energy of the crystal momentum states without changing their wavefunctions, the band topology will stay the same as the Landau level. In particular the Chern number $C = 1$. 

Thus at $U_0 = 0$, at a filling $0 < \nu < 1$, we get a free Fermi sea for fermions, or a condensate for bosons. For small $U_0$, this will evolve into a weakly coupled Fermi liquid (for fermions) or a superfluid (for bosons). Remarkably despite the Hamiltonian $H$ having no explicit kinetic energy,  the purely potential term produces a kinetic energy and enables a conducting phase within the lowest Landau level. 

\begin{figure}
    \centering
    \includegraphics[width=0.9\linewidth]{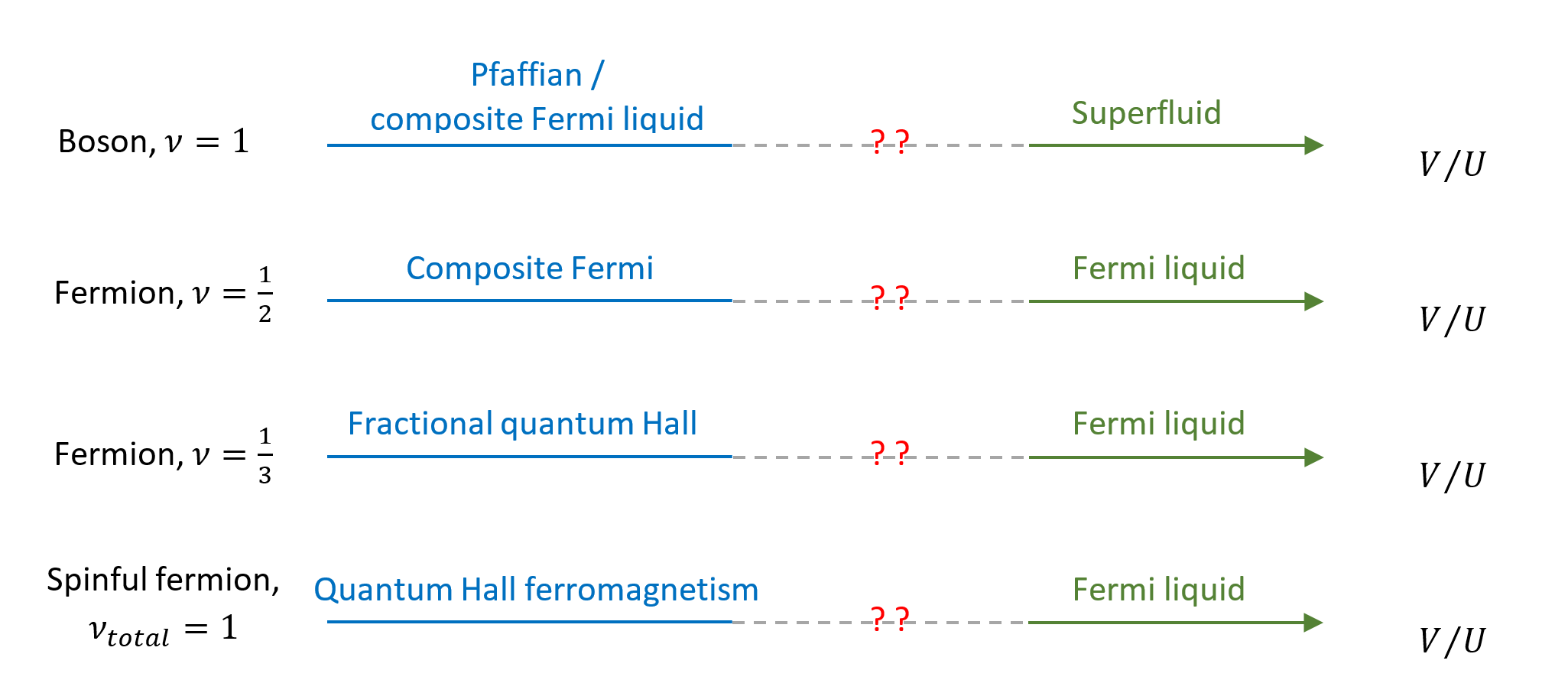}
    \caption{Some example phase diagrams.}
    \label{phasediagram}
\end{figure}

The simple models in Eqn.~\ref{Hamiltonian} thus describe the evolution from the fractional quantum Hall regime to a weakly interacting conducting phase. For some examples of the phases of the model in the two extreme limits of $\frac{V_0}{U_0}$ see Fig.~\ref{phasediagram}. 

This kind of question can in principle be posed for any Chern band. However the advantage of the present models is that we know for sure that in the strong coupling limit we reach the fractional quantum hall state.   For $V_0/U_0$ small enough that the fractional quantum Hall state is stabilized we can regard it as a reliable construction of a fractional Chern insulator on which there is an extensive literature (for reviews, see Ref. ~\onlinecite{bergholtz2013topological, parameswaran2013fractional}). 

It is interesting to also consider situations where in the $V_0 = 0$ the quantum Hall regime the ground state is a metallic composite fermi liquid (as famously happens for fermions at $\nu = 1/2$). The composite fermi liquid will essentially be stable to introducing a small $V_0$ but will eventually evolve into the Fermi liquid (or superfluid if we were considering bosons) in the large $V_0$ limit. 

Thus the models in Eqn.~\ref{Hamiltonian} offer a concrete and simple context to study the interplay between interactions, bandwidth and band topology. We will describe methods that enable us to address analytically the phase diagram and other properties in an example. It should also be possible to study the ground state of such models numerically, for instance using DMRG methods, in the future.   There exists some previous work\cite{grushin2015characterization} studying the evolution between the $\nu =1/3$ fractional Chern insulator and a Fermi liquid metal of spinless fermions in the Haldane honeycomb lattice. The models we discuss are simpler (for example, they have constant Berry curvature and quantum metric, and do not involve the extra unoccupied band with opposite Chern number present in the Haldane lattice), and hence may be easier to study and to interpret.  

In what analytic theoretical framework can we study these models? Here we have the difficulty that even in the classic quantum Hall setting ({\em i.e} Eqn.~\ref{Hamiltonian} without the periodic potential) there exists very little microscopic analytic treatment of the physics. Much has been understood by writing down variational ground state wavefunctions for diverse quantum Hall states\cite{jainbook}. This is very powerful in thinking about gapped topological phases of matter but less so in dealing with gapless phases of matter, or in our goal of studying the evolution with a periodic potential. Traditional methods in quantum Hall physics such as flux attachment mean field theories (and the resulting effective field theories) are usually not microscopically faithful to the lowest Landau level restriction, and hence are not of direct value to us. A notable exception is a system of bosons at filling fraction $\nu = 1$.  Here  by using  a representation\cite{Pasquier1998} of the GMP algebra in terms of fermionic partons, Read\cite{read98} discussed a possible composite fermion ground state in a Hartree-Fock approximation, and studied fluctuations in a diagrammatic approach. In our recent work\cite{dong2020noncommutative}, we revisited this theory and obtained a coarse-grained effective field theory for this composite fermi liquid that is faithful to the lowest Landau level restriction. This effective theory  is a {\em non-commutative} field theory of composite fermions at finite density coupled to a fluctuating emergent $U(1)$ gauge field. We showed that an approximate mapping, valid in the limit of long wavelength, low amplitude, gauge fluctuations leads to the familiar Halperin-Lee-Read theory\cite{hlr} but with parameters determined correctly by the interaction strength, and with calculable corrections\footnote{A proposal for a non-commutative field theory for fermions at $\nu = 1/2$ has recently appeared\cite{govcanin2021microscopic}.} .  

\begin{figure}
    \centering
    \includegraphics[width=0.9\linewidth]{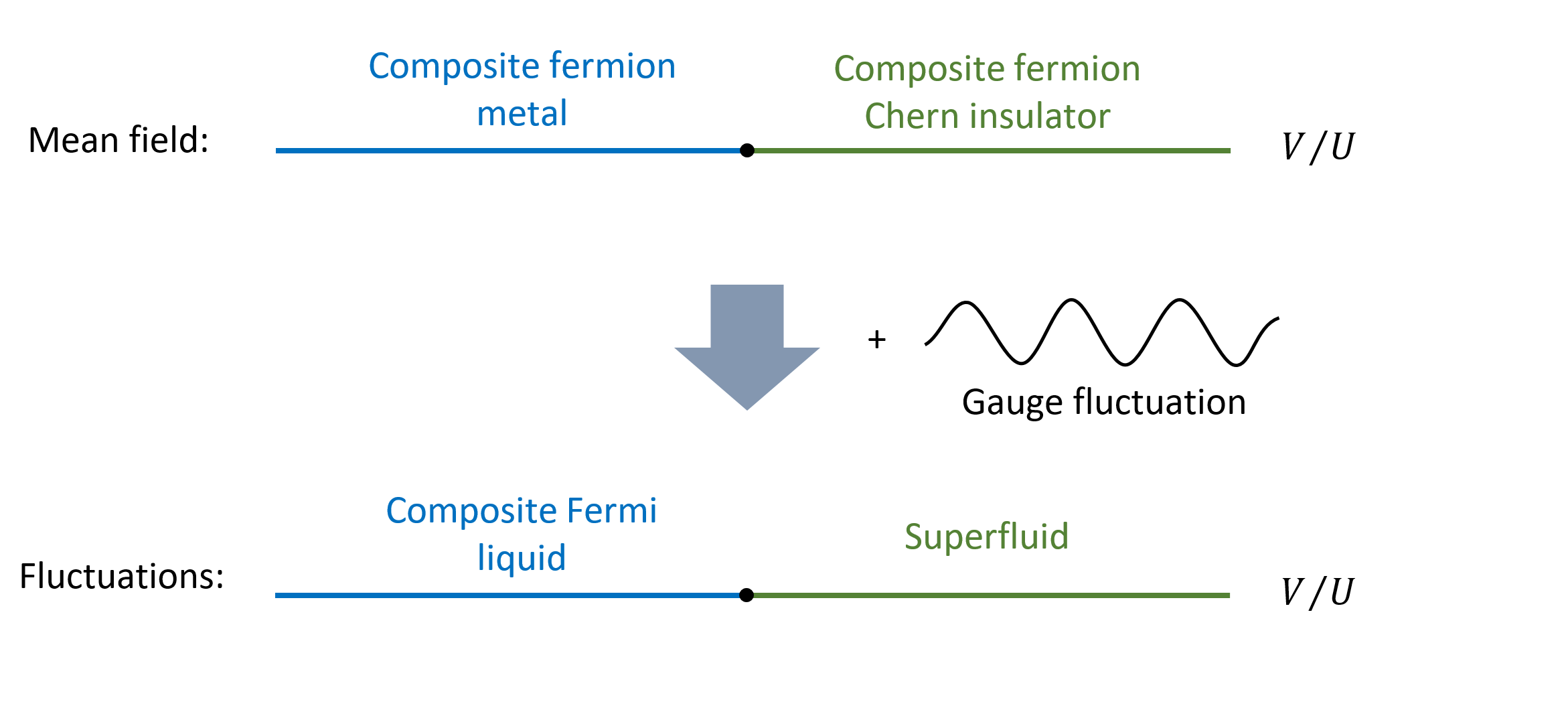}
    \caption{The phase diagram before and after including gauge fluctuation.}
\label{fig3}
\end{figure}

In the rest of this paper we will build on these results and describe the physics of the model in  Eqn.~\ref{Hamiltonian} for bosons at a filling fraction $\nu = 1$. We will show how within the framework of the  fermionic parton description the effect of the periodic potential is readily incorporated.  In the  limit of small $\frac{V_0}{U_0}$ we obtain a composite fermi liquid state\footnote{It is likely that this state is energetically less preferred over a descendant paired state; we will however mostly ignore such paired states. They can be incorporated within our theoretical formalism if needed.} deformed by the presence of the periodic potential. As the density of composite fermions 
is equal to the number of states in the Brillouin zone, the deformed Fermi surface consists of particle and hole pockets of equal area, {\em i.e} it is a composite fermion semi-metal. With increasing $\frac{V_0}{U_0}$, the size of these pockets shrink and eventually there is a phase transition to a composite fermion band insulator. We will show that this state is in fact the superfluid phase of the bosons expected at large $\frac{V_0}{U_0}$. 
Thus both phases and their evolution into each other can be described within a single theoretical framework.   A pictorial depiction of our description is in Fig. ~\ref{fig3}.  

An effective field theory that captures both phases and their evolution is a non-commutative field theory of composite 
fermions in a periodic potential with the action
\be
   {\cal S}[A]  = \int d^2 \v x d\tau~ \overline{c}*D_0 c + \frac{1}{2m^*} |D_i c|^{2} - i a_0\underline{\rho} + \overline{c}(\v x)*V(\v x-\v {\hat{z}}\times \v A(\v x))*c(\v x) 
\ee
Here $c(\v x)$ is the composite fermion field, and $a_\mu$ is a dynamical (non-commutative) $U(1)$ gauge field. $A_\mu$ is a background (non-commutative) $U(1)$ gauge field that couples to the global $U(1)$ currents of the system. $V(\v x)$ is the periodic potential, and takes the form 
\be
V(\v x) = V\left(\cos\left(\frac{2\pi x}{a}\right) + \cos\left(\frac{2\pi y}{a}\right) \right) 
\ee
$a = \sqrt{2\pi}$ is the lattice spacing. The star product that captures the non-commutative structure is defined below, as are the covariant derivatives. The ``non-commutativity' parameter $\Theta = -1$. 
It is understood that the composite fermions are at a non-zero density $\underline{\rho} = \frac{1}{2\pi}$. 
In the vast literature on non-commutative field theories we have not found a discussion of theories with this specific structure, and in particular the crucial periodic potential term. Thus our analysis may also be interesting for the insights it provides into non-commutative quantum field theory.


\section{Review of non-commutative theory for composite fermi liquids}
We begin with a lightning review of the theory of the $\nu = 1$ bosonic composite fermi liquid.  Consider the model in Eqn.~\ref{Hamiltonian} in the absence of any periodic potential, {\em i.e}, in the familiar setting of the continuum lowest Landau level. 
Given a basis set $|m \rangle $ ($m = 1,....., N$) of one-particle states for the Landau level, the many particle Hilbert space of bosons at $\nu = 1$ is defined by the states 
\be
\label{bos_hs}
|\psi\rangle = \Sigma_{\{m_i\}}{a_{m_1,......m_N}} | m_1,........, m_N\rangle 
\ee
with the $a_{m_1,....,m_N}$ symmetric under permutations.   
 
 We use a representation of the GMP algebra in terms of canonical fermion operators $c_{\v k}$ found by Pasquier and Haldane\cite{Pasquier1998}, and developed  by Read\cite{read98}. We write 
 \be
 \label{phrrho}
 \rho_L(\v q) = \int \frac{d^2\v k}{(2\pi)^2}  c^\dagger_{\v k - \v q} c_{\v k} e^{i{\frac{  \v k \times \v q}{2}}}
 \ee
 The fermion operators satisfy the usual anticommutation relations 
 \be
 \{c_{\v k} , c^\dagger_{\v k'}\} = (2\pi)^2\delta^{(2)} (\v k - \v k') 
 \ee
 This is a redundant description, and requires dealing with a constraint 
 \be
 \label{phrcnstr}
  \rho_R (\v q) = \int  \frac{d^2\v k}{(2\pi)^2}  c^\dagger_{\v k - \v q} c_{\v k} e^{-i \frac{ \v k \times \v q}{2}}~= (2\pi)^2\underline{\rho} \delta^{(2)} (\v q) 
  \ee
 Here  $\underline{\rho} = \frac{B}{2\pi } = \frac{1}{2\pi l_B^2}$ is the mean density. 
  The $\rho_R$ also satisfy a GMP algebra but with sign opposite to Eqn.~\ref{gmp}. Furthermore $\rho_R$ commutes with $\rho_L$ at all momenta and hence with the Hamiltonian itself.   Note that the $\v q \rightarrow 0$ limit of Eqn.~\ref{phrrho} implies that the total number of composite fermions equals the number of physical bosons. 
  
Substituting Eqn. \ref{phrrho} into Eqn.~\ref{Hamiltonian} at $V = 0$ gives a four-fermion Hamiltonian which must be solved together with the constraint Eqn.~\ref{phrcnstr} imposed. 
A simple Hartree-Fock approximation that respects translation symmetry seeks a solution where 
\be 
\langle c^\dagger_{\v k} c_{\v k} \rangle \neq 0
\ee
The resulting Hartree-Fock Hamiltonian takes the form 
\be
{\cal H}_{HF} = \int \frac{d^2\v k}{(2\pi)^2} \epsilon_{\v k} c^\dagger_{\v k} c_{\v k} 
\ee
The composite fermions then form a Fermi sea, and we get a mean field description of a composite Fermi liquid.  To treat fluctuations beyond Hartree-Fock, we note that the Hartree-Fock ``order parameter" $c^\dagger_{\v k} c_{\v k} $ does not commute with 
$\rho_R(\v q)$ except at $\v q = 0$. Thus the huge group of gauge transformations generated by $\rho_R$ is broken spontaneously (Higgsed). The important fluctuations are those generated by  $\v q \approx 0$ - these can be thought of as a $U(1)$ gauge field. Thus we should expect to end up with an effective description in terms of a Fermi surface of composite fermions coupled to an emergent dynamical $U(1)$ gauge field.  The precise form of this effective theory  was obtained in Ref.~\cite{dong2020noncommutative} and takes the form of a {\em non-commutative}  field theory with the action 
\be
\label{nccfllag}
{\cal S} = \int d^3x~ \overline{c} * D_0 c + i a_0  \underline \rho +  \frac{1}{2m^*} \overline{D_i c} D_i c \ee
Here the covariant derivatives are defined through 
\be
\label{nccdrv}
D_\mu c = \partial_\mu c - i c*a_\mu - i A_\mu* c
\ee
where $a_\mu$ ($\mu = 0, 1,2$) is a dynamical $U(1)$ gauge field and $A_\mu$ is an external background $U(1)$ gauge field. The composite fermion effective mass $m^*$ is determined by the interaction strength. The composite fermions have a density $\underline{\rho}$. The star product that appears in the action  is defined as follows: given any two fields $f(x)$ and $g(x)$, 
\begin{equation}
    f(x) * g(x) = \lim_{x' \rightarrow x} e^{\frac{i}{2} \Theta \epsilon_{ij}  \partial_i \partial_{j'} } f(x) g(x') 
\end{equation}
Here $\Theta = - l_B^2$ is the ``non-commutativity parameter".  Ref.~\onlinecite{dong2020noncommutative} showed, using a tool known as the Seiberg-Witten map\cite{Seiberg_1999},  that long wavelength low amplitude gauge fluctuations in this non-commutative field theory can be  approximately mapped  to a commutative one and yields the HLR action with calculable corrections.

\section{Composite fermion representation in a periodic potential}
\label{sec:phase diagram}
With this background we take up the problem of studying Eqn.~\ref{Hamiltonian} for bosons at $\nu = 1$.   Compared to the pristine Landau level we have the  extra periodic potential which broadens the Landau level into a Chern band. However as the full Hamiltonian term is written in terms of the density operator $\rho^L$ it may be readily re-expressed using the composite fermion representation of Eqn.~\ref{phrrho}. We will exploit this to analyse the phase diagram.

As explained in our previous work\cite{dong2020noncommutative}, we  rewrite the interaction term as
\be
    H_2 = \int \frac{d^2\v q}{(2\pi)^2} \tilde{U}(\v q)(\rho^L_{\v q}-\rho^R_{\v q})(\rho^L_{-\v q}-\rho^R_{\v q})
\ee
which yields physically sensible results. Similarly, for the one-body potential term we will use
\be
    H_1= \int \frac{d^2\v q}{(2\pi)^2} \tilde{V}(\v q)\rho^L_{-\v q}
\ee
In principle we could add to this a one-body `right' potential term, {\em i.e} a term linear in $\rho^R(-\v q)$ with an arbitrary $\v q$ dependent coefficient. Due to the constraint in Eqn.~\ref{phrcnstr} the Hamiltonian remains the same within the physical Hilbert space. However, we will also eventually introduce the (non-commutative) gauge field $a_\mu$ whose time component $a_0$ couples linearly to $\rho^R$. Any `right' potential can be absorbed into $a_0$, and hence we will not explicitly include any further `right' potential.

The composite fermion Hamiltonian, together with the constraint, can now be approximately treated in a Hartree-Fock approximation. At $V_0 = 0$ this was done in Ref.~\onlinecite{dong2020noncommutative} and yields the composite fermi liquid (or the bosonic Pfaffian if pairing is allowed). Here we will extend the treatment to $V_0 \neq 0$. 

We also specialize to a model where the periodic potential only has harmonics at wave-vectors $\pm \v G_x = \frac{2\pi}{a} (\pm 1,0) $ and $\pm \v G_y = \frac{2\pi}{a} (0,\pm 1)$. In other words 
\be
V(\v q) = V_0 \sum_{s = \pm 1}  \delta^{(2)} (\v q - s\v G_x) + \delta^{2}(\v q - s \v G_y) 
\ee

Let us now discuss the two limits of large and small $U/V$.

\underline{$U\ll V$, weak interaction.}
In terms of the composite fermion, the single particle potential $H_1$ becomes
\be
\label{h1bragg}
    H_1 = V \sum_{\v G}\int \frac{d^2 \v k}{(2\pi)^2}c_{\v k}^\dagger c_{\v k+\v G} e^{\frac{i}{2}\v k \times \v G}
\ee
This periodic potential breaks continuous translational symmetry down to discrete translation symmetry by a lattice vector. 
Eqn.~\ref{h1bragg} can be interpreted as Bragg scattering with a nontrivial pre-factor, which leads to the gap opening at the Brillouin zone boundary for $c$-fermions. 

\underline{$U\gg V$, strong interaction.} In the $V = 0$ limit, at the Hartree-Fock level the interaction induces a composite fermion dispersion that dominates over Bragg scattering, giving the composite fermion a metallic band structure.  As demonstrated in our previous work\cite{dong2020noncommutative},  if we allow composite fermion pairing, then the ground state prefers to have a non-zero pair amplitude in the $l = \pm 1$ channels.  Thus the true Hartree-Fock ground state is a paired state of composite fermion, in line with the result from exact diagonalization\cite{cooper2001quantum}. We note further that for $V=0$,  the Hartree-Fock degeneracy of the  $l=\pm1$ paired states  is guaranteed by the symmetry  of the effective Hamiltonian under an anti-unitary operation on composite fermions that exchanges $\rho^L$ and $\rho^R$ (note that this is not the physical time-reversal since it does not affect the gauge fields). This degeneracy between these two $p$-wave pairing states will be  lifted when gauge fluctuation is taken into account, since the physical problem does not have  such a discrete anti-unitary symmetry.

\section{Band structure of composite fermions}
\label{cfband}

In this section, we will exactly solve the composite fermion band structure close to  the non-interacting limit. In particular we  show that the band has a Chern number which is opposite to that of the LLL that the bosons live in.

First, we note that in general in the presence of the periodic potential right density fluctuations $\rho^R_{\v G}$ will be induced where $\v G$'s are reciprocal lattice vectors. So it is vital that we include a set of Lagrangian multipliers $-ia^0_{\v G}\rho^R_{\v G}$ in the action to enforce the gauge constraint.

Remarkably, in the weak interacting limit $V/U = 0$, the band structure can be solved analytically.
We begin by writing down the general form of Hamiltonian
\be
\label{oneprtcleH}
    H=\sum_{\v G, \v k}V^L_{\v G}c^\dagger_{\v k}c_{\v k+\v G}e^{+\frac{i}{2}\v k\times \v G}+V^R_{\v G}c^\dagger_{\v k}c_{\v k+ \v G}e^{-\frac{i}{2}\v k\times \v G}
\ee
Here, $V_G^L=Ve^{-G^2/2}$ is the strength of periodic potential, while $V^R_{\v G} = \langle a^0_{\v G}\rangle e^{-G^2/2}$ is a variational parameter to be determined so that the composite fermion ground state satisfies the gauge constraint at mean field level, $\langle\rho^R_{\v G}\rangle=0, \text{for } \forall \v G \neq 0$.

The periodic potential, viewed in the momentum basis,  is a scattering between plane wave states $\ket{\v k}$ and $\ket{\v k + \v G_{x,y}}$, or alternatively, a hopping on the momentum space lattice formed by $\v k$-points related through reciprocal lattice vectors $\v G = n\v G_{x} + m\v G_{y}$. Indeed, we have  a momentum space tight-binding model for each $\v k$ in the first Brillouin zone, which will be denoted as $BZ^1$ in the following.
\be
\label{mstb}
    H_{\v k}=\sum_{\v G,\v G'}c^\dagger_{\v k+\v G'}c_{\v k+\v G+\v G'}\bigg(V^L_{\v G}e^{+\frac{i}{2}(\v k+\v G')\times \v G_j}+V^R_{\v G}e^{-\frac{i}{2}(\v k+\v G')\times \v G_j}\bigg)
\ee
where $\v k\in BZ^1$ is the reduced wave vector.
On a side note, in presence of interaction or other terms that lead to single fermion dispersion, there should be an additional trapping potential in the momentum space. But here we first  deal with the non-interacting limit.

\begin{figure}
    \centering
    \includegraphics[width=0.35\linewidth]{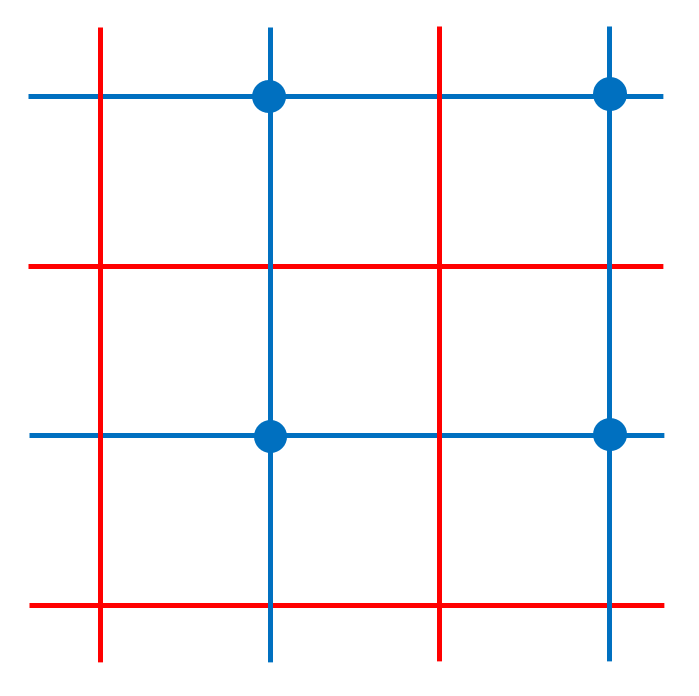}
    \caption{Hopping phases of momentum space tight binding model after gauge transform of Eq.\ref{kgt}. Red and blue lines represent bonds with  $0$ and $\pi$ hopping phase respectively. Each site of this k-space lattice  corresponds to a momentum $\v k+\v G$. The blue dots indicate the sites where a further gauge transform  is performed  to remove all non-zero hopping phases.}
    \label{fig:ktbm}
\end{figure}

At first glance, this momentum space tight binding model is hard to solve since the hopping varies in momentum space. However, we have 
\be
    e^{\pm\frac{i}{2}(\v k+\v G_i)\times \v G_j}=e^{\pm\frac{i}{2}\v k\times \v G_j}(-1)^{(\delta_{ij}+1)}
\ee
(Note that as we are taking $l_B=1$,  $a=\sqrt{2\pi}$ and therefore $G_x=G_y=\sqrt{2\pi}$.)
We emphasis that the additional sign on the right hand side is the same for the form factor of ``left" and ``right" density. Consequently, for the momentum space tight-binding model at each $\v k$ in the first Brillouin zone, the hopping along $x$ has a staggered sign along $y$ direction, and the hopping along $y$ has a staggered sign when translated along $x$. This pattern is shown in Fig.~\ref{fig:ktbm}. The amplitude of hopping is, however, uniform throughout the momentum space. Moreover, note that such a sign pattern can be gauged away and the model becomes symmetric under translation by $\v G_x$ and $\v G_y$. Therefore, the spectrum for each $\v k\in BZ^1$ can be worked out in a straightforward manner. 

Here we pause for a bit and count the degrees of freedom. Consider a finite size system of  size $\sqrt{2\pi}Ml_B\times \sqrt{2\pi}Ml_B$, where the Landau level degeneracy is $N=M^2$, and the set of crystal momenta $(k_x,k_y)$ take values $k_x = \sqrt{2\pi}\frac{ m_x}{M}, k_x = \sqrt{2\pi}\frac{ m_y}{M}$ where $m_x,m_y \in \{0,1,2,...,M-1\}$. The dimension of the single particle Hilbert space for composite fermion is then $N^2$, since $c_{mn}$ has two indices. This suggests that there should be $N^2$ $\v k$-points. It means the number of Brillouin zones in this problem is $N$, since we have $N$ $\v k$-points in the first Brillouin zone. This is also evident from the definition of Fourier transform for a system on the torus
\be
\label{FT}
    c_{\v k} = \sum_{mn}\bra{n}\tau_{-\v k}\ket{m}c_{mn}
\ee
There is an upper-bound for the value of $\v k$, beyond which the magnetic translation operator $\tau_{-\v k}$ translates $\ket{m}$ around the torus by a full cycle, and therefore does not give rise to a new orthogonal basis state. For example, if the system has equal length $\sqrt{2\pi}M$ in $x$ and $y$ directions, then Eqn.~\ref{FT} defines a complete set of plane wave basis $\ket{\v k}=c^\dagger_\v k\ket{0}$ for $k_x, k_y \in \big[-\sqrt{2\pi}M/2,\sqrt{2\pi}M/2\big]$.

Now, we have decoupled the full problem into $N$ momentum space tight-binding models, each having $N$ ``lattice sites". The resulting spectrum has $N$ bands each accommodating $N$ states, and altogether filled by $N$ composite fermions.

In addition, we have a set of variables $\langle a^0_{\v G}\rangle$ that may acquire a finite value to enforce the gauge constraint on right density.

Consider the case with $\langle a^0_{\v G}\rangle =0$ for all non-zero reciprocal lattice vector $\v G$. Then the Hamiltonian becomes
\be
    H=V\sum_{\v G=\pm\v G_x,\pm\v G_y}\rho^L_{\v G}
\ee
In this case, remarkably, the momentum space tight binding models for different $\v k$-points in the first Brillouin zone are related through a gauge transform,
\be
\label{kgt}
    c_{\tilde{\v k}+\v G} \rightarrow \tilde{c}_{\tilde{\v k},\v G} e^{-\frac{i}{2}\tilde{\v k}\times \v G}
\ee
so the spectrum for different $\tilde{\v k}\in BZ^1$ are exactly the same. In other words, the full spectrum of composite fermion is $N$ perfectly flat bands. With $N$ composite fermions, only the bottom band is completely filled. 

The band gap is $\Delta \sim o(V/\sqrt{N})$, and hence  vanishes for $U=0$ when $N\rightarrow \infty$. If we turn on a weak interaction $U\ll V$, at Hartree Fock level this gives a dispersion term for composite fermion $H_U=\sum_\v k\epsilon_\v kc^\dagger_\v k c_\v k$.  The  $\epsilon_{\v k}$ can be approximated by a quadratic dispersion with an effective mass $\epsilon_{\v k}\sim \frac{k^2}{2m^*}$ with $m^*\propto U^{-1}$. In this case, for the momentum space tight-binding model, this dispersion term becomes a harmonic trap potential. So at each $\v k$ the continuous spectrum present at $U = 0$  acquires a gap $\Delta$ for $U \neq 0$. Using the harmonic oscillator representation we get $\Delta\sim (VU)^{1/2}\gg U$.
Thus for non-zero $U$ we have a nearly flat lowest band that is separated from higher energy bands by this  band gap $\Delta$. The composite fermions completely fill the lowest band and (at this mean field level) we get a composite fermion insulator. 

Returning temporarily to $U = 0$, note that the occupied single particle state for each $\v k\in BZ^1$ is the ground state of the corresponding momentum space tight binding model. It follows that (using the $H_{\v k}$ defined in Eqn.~\ref{oneprtcleH} with $V^R = 0$)
\be
\label{eqm}
    H_{\v k,\v G_{x,y}}\ket{\psi}\sim\sum_{\v G'}c_{\v k+\v G'}^{\dagger}c_{\v k+\v G'+\v G_{x,y}} e^{\frac{i}{2}\v k\times \v G_{x,y}}e^{\frac{i}{2}\v G' \times \v G_{x,y}}\ket{\psi}=-\ket{\psi},\qquad \text{for} \; \forall \tilde{\v k} \in BZ^1
\ee
Therefore,
\begin{align}
    \rho^R_{\v G_{x,y}}\ket{\psi}=&\sum_{\v k} c_{\v k}^{\dagger}c_{\v k+\v G_{x,y}} e^{-\frac{i}{2}\v k\times \v G_{x,y}}\ket{\psi} \nonumber\\
    =&\sum_{\v k\in BZ^1, \v G'}{c_{\v k+\v G'}^{\dagger}c_{\v k+\v G'+\v G_{x,y}} e^{-\frac{i}{2}\v k \times \v G_{x,y}}e^{-\frac{i}{2}\v G' \times \v G_{x,y}}\ket{\psi}}
\end{align}
Compare this with Eqn.~\ref{eqm},  note that $e^{i \v G\times \v G'}=1$ for $\forall \v G,\v G'$. Therefore,
\begin{equation}
\label{rhoRexp}
    \rho^R_{\v G_{x,y}}\ket{\psi}=\sum_{\v k\in BZ^1}e^{-i \v k \times \v G_{x,y}}H_{\v k,\v G_{x,y}}     \ket{\psi}=\bigg( \sum_{\v k\in BZ^1}-e^{-i \v k\times \v G_{x,y}}\bigg) \ket{\psi}=0
\end{equation}
So this insulating state of composite fermion satisfies the gauge constraint, and is therefore the ground state we are looking for.

Finally we show  that the filled flat band is topologically nontrivial with a Chern number $C=-1$. To demonstrate this, we define a single particle momentum shift operator which has the same form factor as the right density,
\be
    \tilde{\rho}^R_\v q = \sum_{\v k}\ket{\v k}\bra{\v k+\v q}e^{-\frac{i}{2}\v k\times \v q} = \sum_{\v k\in BZ^1 , \v G}\ket{\v k+\v G}\bra{\v k+\v G+\v q}e^{-\frac{i}{2}(\v k+\v G)\times \v q}
\ee
where $\ket{\v k}=c^\dagger_\v k\ket{0}$ are plane wave single-particle states of composite fermions. Note that we have
\be
\label{Bf}
    \tilde{\rho}^R_\v q\tilde{\rho}^R_{\v q'}\tilde{\rho}^R_{-\v q}\tilde{\rho}^R_{-\v q'} = \sum_\v k \ket{\v k}\bra{\v k} e^{-\frac{i}{2}\v k\times \v q}e^{-\frac{i}{2} (\v k+\v q )\times \v q'}e^{\frac{i}{2}(\v k+\v q +\v q')\times \v q}e^{\frac{i}{2} (\v k+\v q' )\times \v q'}=e^{-i \v q \times \v q'}
\ee
Moreover, this operator transforms the single particle Hamiltonian as
\be
\label{kmt}
\begin{split}
    \tilde{\rho}^R_{-\v q}H^c_{\v k, \v G}\tilde{\rho}^R_{\v q} = &\sum_{\v k',\v G_1}\ket{\v k'+\v G_1}\bra{\v k'+\v G_1-\v q}e^{\frac{i}{2}(\v k'+\v G_1)\times \v q}\\
    &\; \cdot \sum_{\v G,\v G'}V_{\v G}\ket{\v k+\v G'}\bra{\v k+\v G'+\v G}e^{\frac{i}{2}(\v k+\v G')\times \v G}\\
    &\; \cdot \sum_{\v k'',\v G_2}\ket{\v k''+\v G_2}\bra{\v k''+\v G_2+\v q}e^{-\frac{i}{2}(\v k''+\v G_2)\times \v q}\\
    =&\sum_{\v G,\v G'}V_{\v G}\ket{\v k+\v G'+\v q}\bra{\v k+\v G'+\v q+\v G}e^{\frac{i}{2}(\v k+\v q+\v G')\times \v G}\\
    =&H^c_{\v k+\v q, \v G}
\end{split}
\ee
This suggests that the right density is the ``momentum space magnetic translation" operator for composite fermions. To be more specific, we can compare these results with real space magnetic translation for bosons, which is defined as
\be
    \rho_{\v q} = \hat{P}_{LLL} e^{i\v q\cdot \v r} \hat{P}_{LLL} e^{q^2/4} = e^{i\v q\cdot \v R} = \tau(\v z\times \v q \frac{l_B^2}{2\pi})
\ee
where $\v R = \hat{P}_{LLL}\v r \hat{P}_{LLL}$ is the guiding center coordinate operator, which satisfies
\be
    [R_i,R_j]=-i\epsilon_{ij}{l_B}^2
\ee
Utilizing this, we find
\be
\label{Bfr}
    \rho_{\v q}\rho_{\v q'}\rho_{-\v q}\rho_{-\v q'}=\rho_{\v q+\v q'}e^{-\frac{1}{2} [\v q\cdot\v R,\v q'\cdot\v R]} \rho_{-\v q-\v q'}e^{-\frac{1}{2}\v [\v q\cdot\v R,\v q'\cdot\v R]}=e^{i\v q\times \v q'}
\ee
Under this magnetic translation, the Hamiltonian (projected to LLL) transforms as
\be
    \rho_{-\v q}H^b(\v R)\rho_{\v q} = \tau\left(-\v z\times \v q \frac{l_B^2}{2\pi}\right)H^b(\v R)\tau\left(\v z\times \v q \frac{l_B^2}{2\pi}\right) = H^b\left(\v R-\v z\times \v q \frac{l_B^2}{2\pi}\right)
\ee
Since the phase showing up in Eqn.~\ref{Bf} and \ref{Bfr} are opposite in signs, the Berry flux in the first Brillouin zone of the composite fermion should also be opposite of the flux through a magnetic unit cell of the boson, which suggests the Chern number for the composite fermion band to be $-1$.

In Appendix.~\ref{Appendix:Chern} we show an explicit calculation for the Chern number. We will now describe the physical origin of this non-trivial band topology. As we have mentioned, the composite fermion is formed by a boson and a vortex living in opposite LLLs, with the relative position between the two determined by the momentum carried by the composite fermion. Now consider a momentum-space translation of the composite fermion within the lowest band, which is flat in the non-interacting limit. This projection to the lowest band means that the guiding center coordinate of the microscopic boson is  pinned by the periodic potential (so that the energy remains untouched during the translation) and is only allowed to change through a discrete jump by a lattice vector, while the vortex coordinates are free to move around continuously. As a consequence, by varying the $\v G$ index of a composite fermion $c_{\v k+\v G}$, we are really shifting the coordinate of its constituting microscopic boson on a lattice. (Indeed, the fact that the Hamiltonian in non-interacting limit becomes invariant under momentum space lattice translation $\v k\rightarrow \v k+\v G$ mirrors the discrete translational symmetry $\v R\rightarrow \v R+\v a$ of the microscopic boson problem, where $\v a$ is an arbitrary lattice vector.) However, as $\v k$ travels continuously around the Brillouin zone, the boson is not allowed to move, whereas the vortex goes around the magnetic unit cell in the opposite magnetic field. Therefore, the Chern number $C=-1$ of the composite fermion band is directly inherited from the opposite Landau level accommodating the vortices.

\subsection{Hartree-Fock mean-field calculation}
The picture described in Sec.~\ref{sec:phase diagram} is verified by a self-consistent Hartree-Fock mean-field calculation. In the $V=0$ limit, with the single particle potential turned off, we have shown in our earlier work \cite{dong2020noncommutative} that the repulsive interaction leads to attractive pairing in $l=\pm1$ channel, and the pairing appears as a weak instability of the metallic composite fermi liquid. As a consequence, the $U\gg V$ side falls in a $p\pm ip$
\footnote{
At the level of mean field, the $p\pm ip$ are degenerate at $V=0$. This degeneracy is lifted by gauge fluctuation or a finite $V$. In the latter case, it will require a Hartree-Fock calculation with both pairing and density channels to see which one is favored at mean-field level, which is not covered in this paper.}
wave superconducting phase for the $c$-fermion, which is a Pfaffian state in the language of physical bosons.

\begin{figure}[h]
    \centering
    \includegraphics[width=0.95 \textwidth]{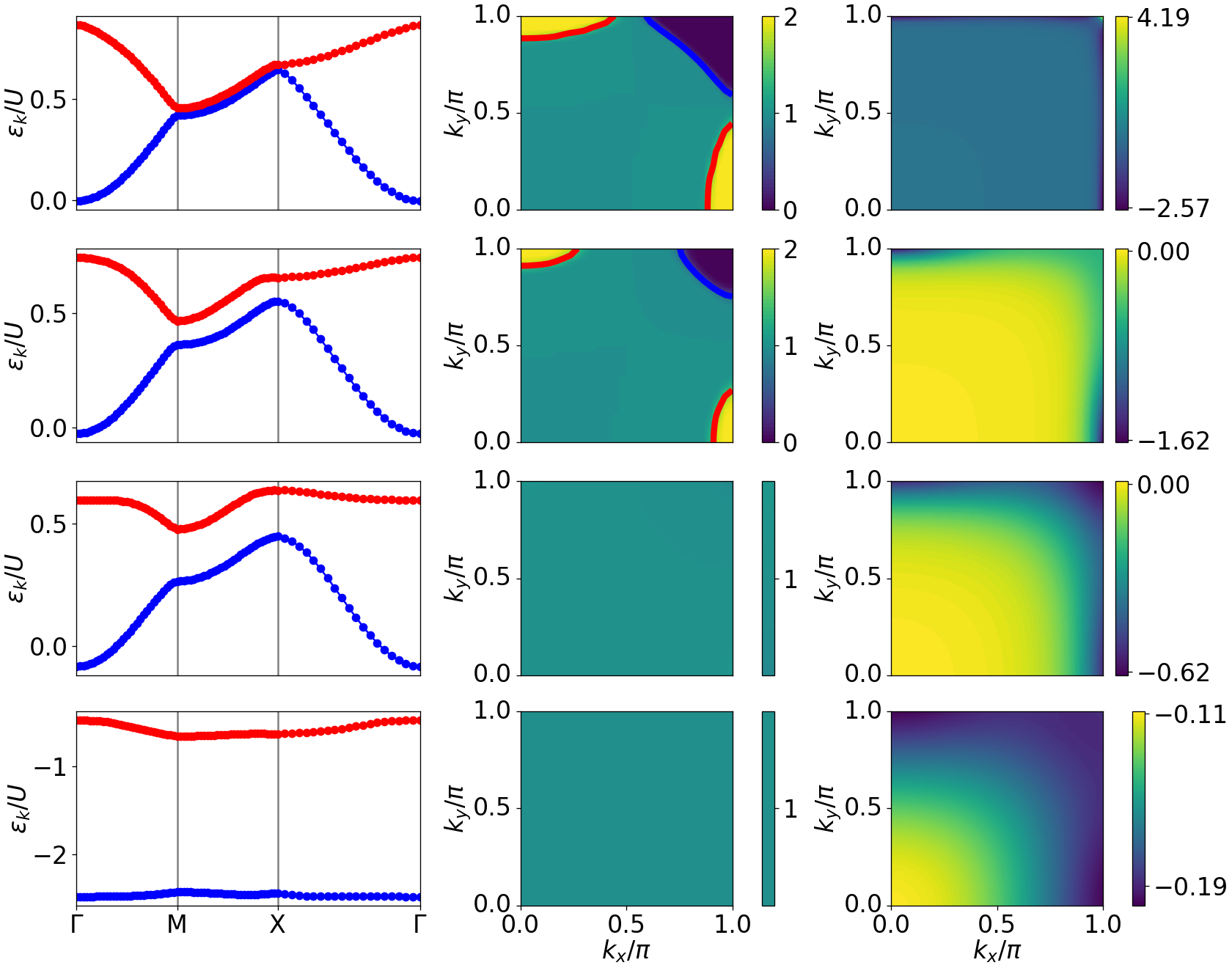}
    \caption{Hartree-Fock mean field calculation: the evolution of (from top to bottom) band structure, Fermi surface, and Berry curvature distribution for composite fermion throughout the phase diagram. From left to right, we plot the results for $V/U=0.1, 0.3, \: 0.5 \:\mathrm{and}\:5.0$. Color scales are shared between plots on the same row, except for the bottom-left plot, because of extreme values contained. Here thanks to the intact four fold rotation symmetry, we reduce the plotted region to a quarter of the first Brillouin zone ($k_x,k_y\in [0,\pi]$). We show in the first line the spectrum on the high symmetry lines, which is indicated by the black lines in the bottom-left panel. In the second row, we plot the momentum distribution of particle number $N_{\v k}=\langle\gamma_{\v k}^\dagger \gamma_{\v k}\rangle$, from which the Fermi surfaces are extracted (blue and red lines for particle and hole pockets, respectively). The metal-insulator transition of composite fermion happens around $V/U=0.45$. For non-zero $V$, no band touching happens. So the band topology is the same for two phases, as is confirmed by a direct computation of Berry curvature.}
    \label{fig:hf}
\end{figure}

To study the evolution with increasing periodic potential, we turn off the Cooper pairing channel in the mean-field and turn up the single particle potential. The numerical results are presented in Fig.~\ref{fig:hf}. In this calculation, $\langle a^0_\v G \rangle$ is varied to reach the saddle point of the action. We find the renormalized right density term to be rather weak in the large $V/U$  phase, namely $V^R/V^L=\langle a^0_\v G\rangle/V<1$, which means the topology of composite fermion always stays the same. In this case, in terms of composite fermions, a phase transition from Chern insulator to fermi liquid is observed around $V/U=0.45$.

In the following, we detail the set up for our Hartree-Fock calculation. We begin with the Hamiltonian
\be
    H = H_1+H_2 = \tilde{U}_{\v q}(\rho^L_{\v q}-\rho^R_{\v q})(\rho^L_{-\v q}-\rho^R_{\v q}) +  \sum_{\v G}\int \frac{d^2 \v k}{(2\pi)^2}\tilde{V}_{\v G}c_{\v k}^\dagger c_{\v k+\v G} e^{\frac{i}{2}\v k \times \v G}+\tilde{V}^R_{\v G}c^\dagger_{\v k}c_{\v k+G}e^{-\frac{i}{2}\v k\times \v G}
\ee
where $\tilde{U}_{\v q} = Ue^{-\frac{q^2}{2}}$ and $\tilde{V}_{\v q} = Ve^{-\frac{q^2}{2}}$, while $V^R_{\v G}$ is a set of parameters to be determined so as to  enforce the  gauge constraint at mean field level.
Now we do the mean field decoupling for the interaction term. For our purpose, we only include the density channel
\be
\label{eq:mfaz_rho}
    \langle c^\dagger_{\v k}c_{\v k'}\rangle = \int_{BZ} \frac{d^2\tilde{\v k}}{2\pi}\sum_{\v G,\v G'} \delta(\v k-\tilde{\v k}-\v G)\delta(\v k'-\tilde{\v k}-\v G')\rho(\tilde{\v k},\v G,\v G')
\ee
We emphasise that we have $\langle c^\dagger_{\v k}c_{\v k+\v G}\rangle \neq 0$.
The mean field Hamiltonian then becomes
\be
\begin{split}
    H_{mf} =& \sum_{\v k, \v G, \v G'}c^\dagger_{\v k+\v G}H(\v k)_{\v G,\v G'}c_{\v k+\v G'} \\
    =&\int_{BZ} \frac{d^2\v k}{2\pi} \sum_{\v G}\bigg(U(1-e^{-\frac{(\v k+\v G)^2}{2}})-\mu\bigg)c^\dagger_{\v k+\v G}c_{\v k+\v G}\\
    &+\int_{BZ} \frac{d^2\v k}{2\pi} \sum_{\v G,\v G'}e^{-\frac{(\v G - \v G')^2}{2}}c^\dagger_{\v k+\v G}c_{\v k+\v G'}\bigg[V_{\v G - \v G'}e^{\frac{i}{2}(\v k+\v G)\times (\v k+\v G')} + V^R_{\v G - \v G'}e^{-\frac{i}{2}(\v k+\v G)\times (\v k+\v G')}\bigg]\\
    &+\int_{BZ} \frac{d^2\v k}{2\pi} \frac{d^2\v k'}{2\pi} \sum_{\v G,\v G',\v g}\bigg[2\tilde{U}_{\v G'-\v G}\lambda(\v k+\v G,\v k+\v G') \lambda(\v k'+\v G'+\v g,\v k'+\v G+\v g)\\
    &+2\tilde{U}_{\v k'-\v k+\v g}\lambda(\v k+\v G,\v k'+\v G+\v g) \lambda(\v k'+\v G'+\v g,\v k+\v G')\bigg]\rho(\v k',\v G'+\v g,\v G+\v g) c^\dagger_{\v k+\v G}c_{\v k+\v G'}
\end{split}
\ee
where $\lambda(\v k,\v k+\v q) = 2i\sin{(\frac{\v k\times \v q}{2})}$ is the form factor in our modified density operator $\rho^L_{\v q}-\rho^R_{\v q}$.

To solve the mean field problem self-consistently, we start with some random set of mean fields $\rho(\v k,\v G,\v G')$, diagonalize $H(\v k)$ to find the eigenmode
\be
    \gamma_{\v k,n} = \sum_{\v G}u(\v k)_{n,\v G}c_{\v k +\v G}
\ee
and the spectrum $\epsilon_{\v k,n}$. Then we always tune the chemical potential to keep $\sum_{\v k, n}\gamma^\dagger_{\v k,n}\gamma_{\v k,n}=N$. In the meantime, the $\langle a^0_{\v G} \rangle$'s, or equivalently $V^R_{\v G}$'s, are also tuned to suppress $\langle \rho^R_{\v G} \rangle  = \sum_{\v k} \langle c^\dagger_{\v k+\v G}c_{\v k+\v G'}\rangle e^{-\frac{i}{2}(\v k+\v G)\times (\v k+\v G')}$. Once these conditions are met, the mean field gets updated to
\be
    \rho(\v k,\v G,\v G') = \langle c^\dagger_{\v k+\v G}c_{\v k+\v G'}\rangle = \sum_{nn'} u(\v k)_{n,\v G} \langle \gamma^\dagger_{\v k,n}\gamma_{\v k,n'}\rangle u(\v k)_{n',\v G'}^*
\ee
This closes the loop and we can iteratively find the self-consistent mean field solution.

Note that we do not include the pairing channel in this calculation. In practice, it is possible that the exact phase diagram may have a unpaired composite Fermi-liquid region sitting in between the bosonic Pfaffian and superfluid phases. In this case, the Bragg scattering $V_{\v G}$ may be just enough to make the pairing interaction no longer attractive in $l=\pm1$ channel while only partially gapping out the composite Fermi sea.

\section{Fluctuations beyond the mean field}
\label{subsec:SF effective theory}
For $V = 0$ the fluctuations beyond the mean field are conveniently described through a non-commutative field theory as discussed in our previous work. There we also showed that for long wavelength low amplitude gauge fluctuations the non-commutative theory can be approximately mapped to a commutative field theory, which takes the same form as the HLR action with some calculable corrections. To handle non-zero $V$ we directly add a periodic potential to the non-commutative field theory. At the mean field level, the action for this non-commutative field theory takes the form
\be
\label{eq:mfaction}
{\cal S}_{mf} = \int d\tau d^2{\v x}~  \bar{c}(\v x, \tau) * \partial_0 c(\v x, \tau)  + \frac{1}{2m^*} \partial_i \bar{c}(\v x, \tau)\partial_i c(\v x, \tau) + \bar{c}(\v x, \tau)* V(\v x) * c(\v x, \tau) 
\ee
It is understood that the composite fermions are at a non-zero density $\underline{\rho}$. Here $V$ is the periodic potential 
\be
\label{eq:Vnc}
V(\v x) = V\left(\cos\left(\frac{2\pi x}{a}\right) + \cos\left(\frac{2\pi y}{a}\right) \right) 
\ee
with $a^2 = 2\pi |\Theta|$. (In the conventions of the previous sections, the non-commutative parameter $\Theta = - \frac{1}{B} \equiv -1$.)
We have simplified the mean field theory by only keeping the external potential that couples to the `left' density.  In particular we ignore the one-body terms that mix the left and right densities discussed in the previous section. They do not affect our discussion of universal aspects of the physics (and further are small in the two extreme limits $U/V \gg 1$ and $U/V \ll 1$). 

To go beyond the mean field we need to include gauge fluctuations $a_\mu$ that couple to the `right' density and currents.   To that end we  replace derivatives $\partial_\mu$ by covariant derivatives $D_\mu$ : 
\be
\label{eq:coderiv}
D_\mu c = \partial_\mu c - i c*a_\mu 
\ee
 For now we have only  included   the internal ({\em i.e} dynamical) gauge field $a_\mu$. Later we discuss how to properly couple external probe gauge fields that couple to the `left' densities and currents. 

The effective action\footnote{Recall that the Hartree-Fock calculation generates other one body potentials of the structure $\bar{c}* V_1 * c* V_2$. Indeed such a term is allowed by the symmetries of the model. Nevertheless, we have not included it in Eqn. ~\ref{eq:effftV} so as to obtain a minimal field theory that captures universal aspects of the phase diagram. Quantitatively this extra term will have small effects: it will clearly be small for small $V_0$. For large $V_0$, as we argued in Sec.~\ref{cfband}, the composite fermion band gap is $\Delta\sim O(\sqrt{UV})$, while this  extra potential leads to a renormalization of the dispersion with   $V_1,V_2\sim O(U)\ll \Delta$.}
 that captures fluctuations beyond the mean field then becomes
\be
\label{eq:effftV}
{\cal S} = \int d\tau d^2{\v x}~  \bar{c}(\v x, \tau) * D_0 c(\v x, \tau)   + i a_0 \underline{\rho}  + \frac{1}{2m^*} \overline{D_i c} D_i c   + \bar{c}(\v x, \tau)* V(\v x) * c(\v x, \tau) 
\ee

This action is manifestly invariant under non-commutative `right' $U(1)$ gauge transformations. Under renormalization we therefore will only generate terms that preserve this gauge invariance. 
Let us consider the phase diagram as the strength of the periodic potential is varied. At zero periodic potential this action describes the composite fermi liquid state in the continuum Landau level. For long wavelength gauge fluctuations we can use the Seiberg-Witten map to obtain an approximate commutative field theory (the HLR theory). 
A weak periodic potential will provide very little modification of this state (except to reconstruct the composite  fermi surface where it intersects the Brillouin zone boundary). 

We therefore turn to the opposite limit of weak interaction $V\gg U$, where the composite fermions acquire an insulating band structure. We will see below that upon including the gauge fluctuations, the composite fermion band insulator correctly describes the expected superfluid of the microscopic bosons. 
However as the periodic potential varies on the scale of the magnetic length we need to treat it more accurately. Accordingly to discuss the fate of the large-$V$ limit (when the composite fermions form a band insulator) we will not use the Seiberg-Witten map. 

In the presence of a strong periodic potential, the composite fermion is gapped and can be integrated out.  The effective action for the $a_\mu$ fields that results is severely constrained by the non-commutative gauge invariance of the theory. To leading order in a derivative expansion this action must take the form 
\be
\label{seff} 
{\cal S}[a_\mu]  = {\cal S}_{CS}[a_\mu] + {\cal S}_{Max}[a_\mu]
\ee
The first term is the non-commutative Chern-Simons action: 
\be
\label{eq:scs}
{\cal S}_{CS} = \int d^2{\v x} d\tau \frac{k}{4\pi} \epsilon_{\mu\nu\lambda} \left( a_\mu *\partial_\nu a_\lambda + \frac{2i}{3} a_\mu * a_\nu * a_\lambda \right) 
\ee
The level $k$ is necessarily quantized to be an integer. The second term in Eqn.~\ref{seff} is the non-commutative Maxwell action\footnote{Strictly speaking as we are dealing with a non-relativistic system, we should have different coefficients for the electric and magnetic field terms in the Maxwell action. For notational convenience we will not write this explicitly. }  
\be
\label{smax}
{\cal S}_{Max} = \frac{K}{2} \int d^2{\v x} d\tau f_{\mu \nu} * f_{\mu \nu}
\ee
with the non-commutative field strength  $f_{\mu \nu} = \partial_\mu a_\nu - \partial_\nu a_\mu + i (a_\mu * a_\nu - a_\nu * a_\mu)$. $K$ is a positive constant that will be determined by microscopic parameters. 

The quantized coefficient $k$ can be found through explicit calculation by integrating out the fermions. Interestingly we find that $k = 0$ so that the long wavelength effective action is described purely by the Maxwell theory. Details of the computation are in Appendix \ref{Appendix:CS}. The result $k = 0$ is at first sight surprising as we have shown that the filled band of the composite fermions has a non-zero Chern number. However the coupling to the internal gauge field $a_\mu$ when written out in $\v k$-space has a structure that distinguishes it from an ordinary gauge field that shifts all momenta from $\v k$ to $\v k + \v a$. In particular though the periodic potential term has momentum dependence it does not couple to the internal gauge field. 

To  understand the physics described by the Maxwell action, let us now ask how an external background gauge field $A$ couples to the action in Eqn.~\ref{eq:effftV}. The naive procedure is to couple $A$ minimally so as to have invariance under non-commutative `left' $U(1)$ gauge transformations. For the derivative terms such gauge invariance is readily achieved by generalizing the covariant derivative defined in Eqn.~\ref{eq:coderiv} to 
\be
\label{eq:coderivA}
D_\mu c = \partial_\mu c - i c*a_\mu - i A_\mu * c 
\ee
However the periodic potential term is not invariant under `left' $U(1)$ gauge transformations, and hence will need to also be modified.  Below we will focus on a long wavelength gauge field $A$ and obtain the correct modification of the periodic potential term to leading order in $q^2|\Theta|$ (where $q$ is the momentum of $A$).  Recall that the effect of a left gauge transform  
\be
\begin{split}
   c &\rightarrow e^{i\theta}*c\\
    A_\mu &\rightarrow A_\mu +\partial_\mu \theta + i[\theta,A_\mu]_*
\end{split}
\ee
with $\theta = \v k\cdot \v x$ is nothing but a translation of the non-commutative coordinate of the fermion by $\v z\times \v k$, combined with, in the long-wavelength limit of gauge fluctuation, a constant shift to the $\v A$ by $\v k$. Inspired by this observation, we can modify  the potential term to make the full action invariant under   long-wavelength `left' gauge transformations. The resulting action  takes the form 
\be
\label{eq:S}
   {\cal S}[A]  = \int d^2\v x d\tau ~\overline{c}*D_0 c + \frac{1}{2m^*} |D_i c|^{2} - i a_0\underline{\rho} + \overline{c}(\v x)*V(\v x-\v {\hat{z}}\times \v A(\v x))*c(\v x) 
\ee

In Appendix~\ref{Appendix:CS} we explicitly integrate out the fermions and obtain the induced effective action in the presence of the background gauge field $A$. Strictly speaking we should obtain an effective action that is invariant under arbitrary non-commutative gauge transformations of both $a_\mu$ and $A_\mu$. We will however be satisfied with an expansion of the action in powers of $A, a$ and their gradients, as all we need is the leading order response of the system to a long wavelength background gauge field.  To leading order in this expansion we find  
\begin{eqnarray}
\label{eq:seff_aA} 
{\cal S} & = &   {\cal S}_0[a, A] + {\cal S}_{Max}[a] \\
{\cal S}_0[a,A] & = & \int d^2{\v x} d\tau  -i A_0(\v x) \bar{\rho}
     -\frac{1}{2\pi} \epsilon_{\mu\nu\rho} A_\mu \partial_\nu A_\rho -\frac{1}{2\pi} \epsilon_{\mu\nu\rho} A_\mu \partial_\nu a_\rho
\end{eqnarray} 

Note that the action in Eq.\ref{eq:seff_aA} is gauge invariant under  long-wavelength gauge transformations. It is however not gauge invariant under arbitrary non-commutative gauge transformations. This is due to us not retaining higher powers of $A,a$ and momentum $q$. Based on  this truncated long wavelength action, can we guess the form of a more general action that is invariant under non-commutative gauge transformations?   The self Chern-Simons term involving $A$ can clearly be the remnant of the fully gauge invariant non-commutative Chern-Simons term in the long-wavelength limit. However the effective action above also involves a mutual Chern-Simons term between $a$ and $A$. We have not been able to guess the form of a non-commutative mutual Chern-Simons action for $a,A$ that reduces to the usual form above in the long wavelength limit, and leave this as an interesting excercise for the future. 


Note that since $A$ only has low momentum fluctuations, in the mixed Chern-Simons term only the long wavelength fluctuations of $a$ contribute. To expose the physics of this effective action we now use the Seiberg-Witten map to obtain a commutative effective field theory for long wavelength fluctuations of both gauge fields. 
After the Seiberg-Witten map, the first term contributes another commutative Chern-Simons term for $A$: 
\be
\label{S1}
    {\cal S}_{1}[\hat{A},\hat{a}] = \int d^2{\v x} d\tau \frac{1}{4\pi} \epsilon_{\mu\nu\rho} \hat{A}_\mu \partial_\nu \hat{A}_\rho
\ee
Eventually, combining Eq.\ref{S1} and \ref{eq:seff_aA}, we find the full commutative long-wavelength action
\be
\label{S}
    {\cal S}[\hat{A},\hat{a}] = \int d^2{\v x}d\tau -\frac{1}{4\pi} \epsilon_{\mu\nu\rho} \hat{A}_\mu \partial_\nu \hat{A}_\rho -\frac{1}{2\pi} \epsilon_{\mu\nu\rho} \hat{A}_\mu \partial_\nu \hat{a}_\rho + \frac{K}{2} \left(\epsilon_{\mu \nu \lambda} \partial_\nu \hat{a}_\lambda\right)^2 
\ee
Integrating over dynamical gauge field $\hat{a}$, we find $d\hat{A} \approx 0$ at long wavelengths, which  correctly describes a superfluid phase of the microscopic bosons.

\section{Spinful bosons at $\nu_T=1$}
In this section we briefly discuss the fate of spin-$1/2$ bosons at a total filling $\nu_T = 1$ in the LLL, and in the presence of a periodic potential.  For strong interaction ($U\gg V$), we expect to get a composite fermi liquid of spin-$1/2$ composite fermions, with a possible instability toward a paired quantum Hall state. In the opposite limit of $U \ll V$, the periodic potential will give the bosons a dispersion, and the bosons will condense. The resulting state is a ferromagnetic superfluid. 

Both limits can be understood within the LLL composite fermion description introduced in our earlier work (Ref. ~\onlinecite{dong2020noncommutative}).
For $U \gg V$ we showed there that the mean field ground state is indeed a spinful composite Fermi liquid.  Note now that, when the periodic potential is turned on, the composite fermion Fermi surface sits entirely within the first Brillouin zone and does not intersect the zone boundary. Consequently it is only mildly distorted by the periodic potential. In particular, unlike the spinless case discussed in the bulk of the paper, it does not transform into a composite fermion semi-metal.  A further difference with the spinless case is that the pairing in $l=1$ channel is no longer attractive. Instead, at mean-field level there is an instability toward s-wave pairing \cite{dong2020noncommutative}. 

In the opposite limit ($U\ll V$),  in the composite fermion picture, everything exactly follows our discussion in Sec.~\ref{cfband} -- the spinful  composite fermions fill a nearly flat band with Chern number $C=-1$.  Interactions will then produce a ferromagnetic Chern insulator where only spin species of composite fermion is present and fully fills the band.  The effect of gauge fluctuation for this case is then exactly the same as that of the spinless version, namely as Eqn.~\ref{S}, and this state is a ferromagnetic superfluid of the bosons. 

In contrast to the spinless case, the evolution between these two limits is not straightforward to determine. A simple guess is that with increasing $V/U$, the spin singlet composite fermi liquid first undergoes a Stoner transition to partial spin polarization, which eventually gives way through a second transition to a fully polarized ferromagnetic composite fermi liquid. From that point on, the system essentially evolves in the same manner as the spinless composite fermions discussed in previous sections.

\section{Discussion} 
In this paper we  introduced and studied simple models of  strongly  correlated quantum particles  in a partially filled topological band. We made analytic progress for one specific case (bosons at $\nu = 1$ in the LLL, and in the presence of a periodic potential that gives a finite bandwidth to the LLL).
The models defined in Eqn. ~\ref{Hamiltonian} are well suited to studying the evolution between quantum Hall physics and that of weakly interacting conducting phases for particles at any filling.  To analytically attack cases other than the ones studied here, it will first be necessary to develop a LLL theory of the basic quantum Hall regime ({\em i.e} even without a periodic potential). Given such a theory it will be straightforward to include the effects of the periodic potential. 
We emphasize that it is not sufficient to just know the long wavelength topological field theory (for a gapped quantum hall state) for this purpose. Rather we need a theory that knows about the density operator at scales of order the magnetic length. Even then it will not necessarily be the case that the weak and strong coupling limits can be accessed within the same framework. 

A useful analogy is with the familiar lattice Hubbard model: though the weak and strong coupling limits are often understood, the evolution between them is a challenging problem in condensed matter physics. Despite lack of analytic progress the Hubbard model provides a guide for the basic lattice strong correlation problem, and, in some cases, can be studied numerically. We hope that the models defined here play a similar role for strong correlations in a topological band. 

On the numerical front, these models can be studied with DMRG methods. Variational wavefunctions that have been so successful  in the quantum hall regime may also lend themselves to incorporating the effect of the periodic potential, and may provide useful insight. 

Finally,   our results on the non-commutative field theory  suggest that there is much interesting physics in the presence of a periodic potentail in such theories which may be valuable to pursue.

\section{Acknowledgement}
We thank Adrian Po, Ya-Hui Zhang, and particularly Hart Goldman for many stimulating discussions.   This work was supported by NSF grant DMR-1911666,
and partially through a Simons Investigator Award from
the Simons Foundation to Senthil Todadri. This work was also partly supported by the Simons Collaboration on Ultra-Quantum Matter, which is a grant from the Simons Foundation (651440, TS).

\appendix
\counterwithin{figure}{section}
\section{Chern number of composite fermion band}
\label{Appendix:Chern}
Here we explicitly calculate the Chern number composite fermion band. To do this, we first obtain the Bloch vector $\ket{u(\v k)}$ for $\v k\in BZ^1$, whose entries are $u_{\v G}(\v k)$, where $\v G's$ are Bragg vectors. In particular, according to Bloch's theorem,
\be
\label{eq:bloch}
    \ket{\psi_{n,\v k}} = \sum_{\v G} u_{n,\v G}(\v k) \ket{\v k+\v G} 
\ee
where $\ket{\psi_{n,\v k}}$ is the wavefunction that diagonalizes the Hamiltonian, and $\ket{\v k}$ is the plain wave state. Throughout this appendix we will leave out the band index $n$ for the sake of notational simplicity, since the only band of our interest is the lowest band $(n=0)$, which is fully occupied.
The Bloch states are defined as
\be
    \ket{u_{\v k}} = \sum_{\v G} u_{n,\v G}(\v k) \ket{\v G} 
\ee
In Sec.~\ref{cfband} we discussed the ``magnetic translation" in momentum space. Particularly, using Eqn.~\ref{kmt}, we generate the Bloch state at $\v k+\delta \v q$ from the solution at $\v k$ using the momentum space magnetic translation operator $\tilde{\rho}^R_{-\delta \v q}$
\be
\label{magtr}
    \ket{u_{\v k+\delta\v q}} = \tilde{\rho}^R_{-\delta \v q}\ket{u_{\v k}}
\ee
So the components of the two Bloch state vectors are related through
\be
    u_{\v k+\delta \v q,\v G} =  e^{\frac{i}{2}(\v k+\v G)\times \delta \v q}u_{\v k,\v G}
\ee
As a gauge choice for Bloch vector at every momentum in Brillouin zone, we generate all $u_{\v G}(\v k)$'s from $u_{\v G}(0)$, namely,
\be
\label{A1}
    u_{\v G}(\v k) = e^{\frac{i}{2}\v G\times \v k} u_{\v G}(0) = e^{\frac{i}{2}\v G\times \v k} u^0_{\v G}
\ee
The Berry connection
\be
\label{A2}
    A_j(\v k) = -i\bra{u_\v k}\partial_{k_j}\ket{u_\v k}=\frac{1}{2}\sum_{\v G}|u^0_{\v G}|^2G_i\epsilon_{ij}
\ee
At a glance, this is a constant, which seems to suggest the Berry's curvature $b(\v k)=\nabla \times \v A(k)=0$. But this should not be the case, because even from Eqn.~\ref{A2}, we can smell the non-trivial band topology by rewriting the right-hand-side as $\frac{1}{2}\langle \v z\times \v G\rangle$, and further noting that $\v A(\v k)$ is identified as the polarization of composite fermion modulo the lattice. Therefore, this equation already suggests an equivalence between polarization and the momentum rotated by 90 degrees -- just like what we would expect in a Landau level.

To reveal the topology, first note that the summation in Eqn.~\ref{A2} is not convergent. Therefore the result is ambiguous in the sense that it depends on the UV cutoff. Meanwhile, in the physical problem we are interested in, there is a natural UV cutoff. In particular, recall that technically the Chern number is not well-defined at $U=0$, where the band gap vanishes. This is remedied by introducing a small $U$, as we have discussed in Sec.~\ref{cfband}. By doing so, the Bloch function $u_\v G$, instead of being a plane wave on the lattice spanned by $\v G$, now gets trapped by the $U$ term , and becomes localized around $\v G=0$.

In this case, the Bloch function is $k$ dependent. Eqn.~\ref{A2} becomes
\be
\label{A3}
    A_j(\v k) = -i\bra{u_\v k}\partial_{k_j}\ket{u_\v k}=\frac{1}{2}\sum_{\v G}|u^0_{\v G}(\v k)|^2G_i\epsilon_{ij}
\ee

To be more concrete, we treat the $U$ term within the effective mass approximation $H_U \sim \frac{k^2}{2m^*}c^\dagger_kc_k\sim \frac{Uk^2}{(2\pi)^2}c^\dagger_kc_k$. In this case, we can approximate the Bloch function as
\be
\label{Gaus}
    u_\v G(\v k)|_{U}\sim e^{-\frac{(\v k+\v G)^2}{2\sigma^2}}u_\v G(\v k)|_{U=0} 
\ee
where $\sigma\sim \frac{2\pi V}{U}$ is the size of the Gaussian wave function that solves the harmonic oscillator problem. This approximation works because for $U\ll V$, we have $\sigma \gg 2\pi$, which means the discreteness of lattice $\{\v G\}$ does not show up at the length scale of $\sigma$.

Combine Eqn.~\ref{Gaus} and \ref{A1}, we find at small $U$, the Bloch vector is approximated by
\be
    u_\v G(\v k) \sim u^0_\v G e^{-\frac{(\v k+\v G)^2}{2\sigma^2}}e^{\frac{i}{2}\v G\times \v k}
\ee
Note that since $\sigma \gg 2\pi$, the phase factor $e^{\frac{i}{2}\v G\times \v k}$ varies much faster than the Gaussian profile $e^{-\frac{(\v k+\v G)^2}{2\sigma^2}}$, and that this is always true as long as $U$ is small. The detailed form of the $U$ term is not relevant. Therefore, the leading contribution to the Berry curvature comes from the term with derivatives acting on the phase factor, which gives
\be
\label{A5}
    A_j(\v k) = -i\bra{u_\v k}\partial_{k_j}\ket{u_\v k}=\frac{1}{2}\sum_{\v G}|u_{\v G}(\v k)|^2G_i\epsilon_{ij} 
\ee
which can be rewritten as
\be
\label{A6}
    A_j(\v k) =\frac{1}{2}\sum_{\v G}|u_{\v G}(\v k)|^2(k_i+G_i)\epsilon_{ij}-\frac{1}{2}k_i\epsilon_{ij}
\ee
where we have used the normalization condition $\sum_{\v G}|u_{\v G}(\v k)|^2=1$.
The first term is periodic in the Brillouin zone. We will denote $\frac{1}{2}\sum_{\v G}|u_{\v G}(\v k)|^2(k_i+G_i) = \alpha(\v k)_i$. Therefore, the Berry flux through Brillouin zone is
\be
\begin{split}
    \Phi_B &= \int d^2\v k b(\v k) = \int d^2\v k \epsilon_{ij}\partial_{k_i}A_j(\v k)\\
    &= \int d^2\v k \epsilon_{ij}\epsilon_{lj}\partial_{i} \alpha_l(\v k) -  \frac{\epsilon_{ij}\epsilon_{lj}}{2}\partial_{i}k_l\\
    &=-\frac{\epsilon_{ij}\epsilon_{ij}}{2} \int d^2\v k =-2\pi
\end{split}
\ee

\section{Numerical calculation of Chern number}
\label{appendix:numerical}
In this section we detail the numerical method to extract Berry curvature. In our self-consistent mean field calculation, we keep up to 21 Brillouin zones. Using the converged mean fields, we sample 30 by 30 $k$-points in the first Brillouin zone and diagonalize the mean field Hamiltonian $H_{\v k}$ for each $\v k$ sampled. As a result we obtain the Bloch vector of bottom band $u_{\v G}({\v k})$ defined in Eqn.~\ref{eq:bloch}. 
To obtain Berry curvature, we use
\be
    B(\v k) = \nabla \times \v A(\v k) = -i\epsilon_{ij} \braket{\partial_i \psi_\v k|\partial_j \psi_\v k} = \sum_G -i\epsilon_{ij} \partial_i u_{\v G}(\v k)^* \partial_j u_{\v G}(\v k)
\ee
which is approximately computed by
\be
    \frac{2}{\delta k_x\delta k_y}\text{Im} \sum_\v G \bigg(u_{\v G}(\v k+\delta \v k_x)^*-u_{\v G}(\v k)^*\bigg)\bigg(u_{\v G}(\v k+\delta \v k_y)-u_{\v G}(\v k)\bigg)
\ee
where $\delta k_{x,y}$ is the momentum increment of our sampling, in our case it is set to be $\frac{2\pi}{30}$. The total Berry flux through the first Brillouin zone is however found not to be a quantized number. This is due to the cutoff for large $k$. In fact, the truncated wavefunction $\ket{\psi_{\v k}}$ is not even periodic across the edges of Brillouin zone. But note that the error caused by cutoff becomes weaker as we reduce $V/U$, since the wavefunction $u_{\v G}(\v k)$ gets more localized in momentum space at small $\v G$'s.

\begin{figure}[h]
    \centering    
    \includegraphics[width=0.5 \textwidth]{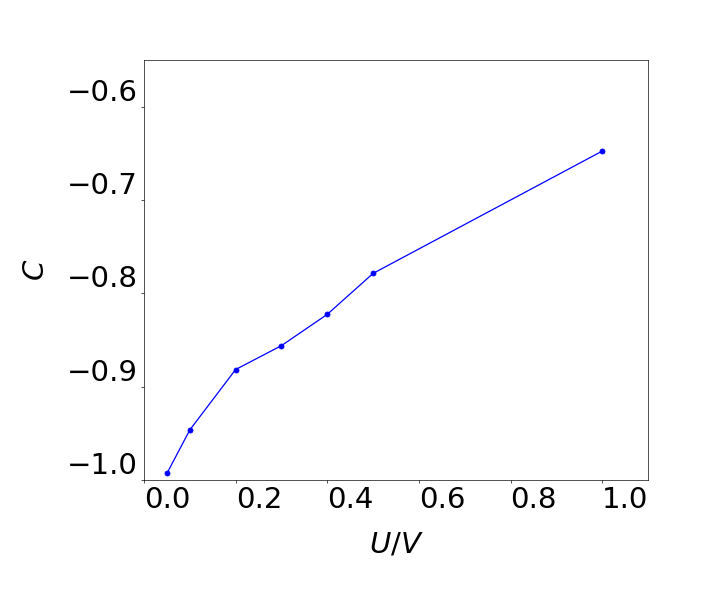}
    \caption{The evolution of measured Chern number as a function of $U/V$. $C\rightarrow -1$ as $U/V\rightarrow 0$}
    \label{fig:C-U}
\end{figure}

Here we show in Fig.~\ref{fig:C-U} the total Berry flux we get using the method and accuracy parameters described above, as a function of $V/U$. Note that even though we need a finite $V/U$ to gap out the composite fermi sea, the direct band gap is opened as long as an infinitesimal $V$ is turned on so that the Chern number for lowest fermion band can be computed even within the composite fermi liquid phase. Indeed, we find the total Berry flux approaches $\frac{\phi_B}{2\pi}=-1$ as $V/U\rightarrow 0$. Here one may notice a small deviation from quantization at small $V/U$. We argue that this is caused by the finite sampling of $k$ points within the Brillouin zone, since for small $V/U$, the Berry curvature is strongly peaked around the edge of BZ, where band gap is opened. In this case a discrete summation no longer approximates the momentum space integral well.

\section{Chern-Simons terms from fermion path integral}
\label{Appendix:CS}
Here we explicitly determine the coefficient of Chern Simons terms of the CF band insulator by integrating out the composite fermions. We start with the non-commutative gauge theory
\be
    {\cal S}[c,\bar{c},a,A] = \int d^3x \: \bar{c}D_0c + ia_0\bar{\rho} + \frac{1}{2m^*} |D_i c|^2 + \overline{c}(\v x)*V(\v x-\v {\hat{z}}\times \v A(\v x))*c(\v x)
\ee
where covariant derivatives are defined as
\be
    D_\mu c = \partial_\mu c + i c * a_\mu + i A_\mu * c
\ee
Eventually we want to get an effective action ${\cal S}_{ind}$ by integrating out the fermions:
\be
    e^{-{\cal S}_{ind}[a,A]} = \int {\cal D}\bar{c}{\cal D}c \: e^{-S[c,\bar{c},a,A]}
\ee

\subsection{Self Chern-Simons term for internal gauge field}
To begin with, for now we restrict ourselves to $A=0$ and examine the internal gauge field response arising from the path integral. As we will show later, terms involving $A$ can be obtained in a similar way. Due to gauge invariance, we expect the following result at long wavelength and to leading order in powers of $a$
\be
    {\cal S}_{ind}[a] = {\cal S}_{CS}[a] + {\cal S}_{Maxwell}
\ee
where the non-commutative Chern Simons term is
\be
    {\cal S}_{CS}= \frac{k}{4\pi}\int d^3x \epsilon_{\mu\nu\rho} a_\mu \partial_\nu a_\rho + \frac{2}{3} a * a * a
\ee
For our purpose to get the Chern Simons level $k$, we only need to consider (in Fourier space) the term $\epsilon_{\mu\nu\rho} q_\mu a_\nu(q) a_\rho(-q)$, where $q_\mu=(\omega,\v q)$ is the 3-momentum. In particular we will look for the coefficient of $\omega a_x(\omega,0) a_y(-\omega,0)$. 

The mean field Hamiltonian is 
\be
\label{H0}
    H^{(0)}_{MF} = H_0+H_V = \sum_{\v k,\v G} \epsilon^0_{\v k+\v G} c^\dagger_{\v k+\v G} c_{\v k+\v G} + \sum_{\v k, \v G, \v Q} V_\v Q c^\dagger_{\v k+\v G+\v Q} c_{\v k+\v G}e^{\frac{i}{2}(\v k+\v G)\times\v Q} 
\ee
where $\epsilon^0_\v k = \frac{\v k^2}{2m^*}$. The gauge fields  $a_x, a_y$ are coupled through
\be
\label{H12}
    H^{(1)}_{MF}+H^{(2)}_{MF} = \sum_{\v q,\v k,\v G}\frac{k_i+G_i+q_i/2}{m^*}a^i_{\v q}e^{\frac{i}{2}(\v k+\v G) \times \v q}c^\dagger_{\v k+\v G+\v q}c_{\v k+\v G} + \frac{1}{2m^*}a^i_{\v q}a^i_{-\v q}c^\dagger_\v kc_\v k
\ee
To extract the coefficent of $a_x a_y$, we can ignore the second term in Eqn.~\ref{H12}, which is diagonal.
\begin{figure}[h]
\label{bubble}
    \centering    
    \includegraphics[width=0.7 \textwidth]{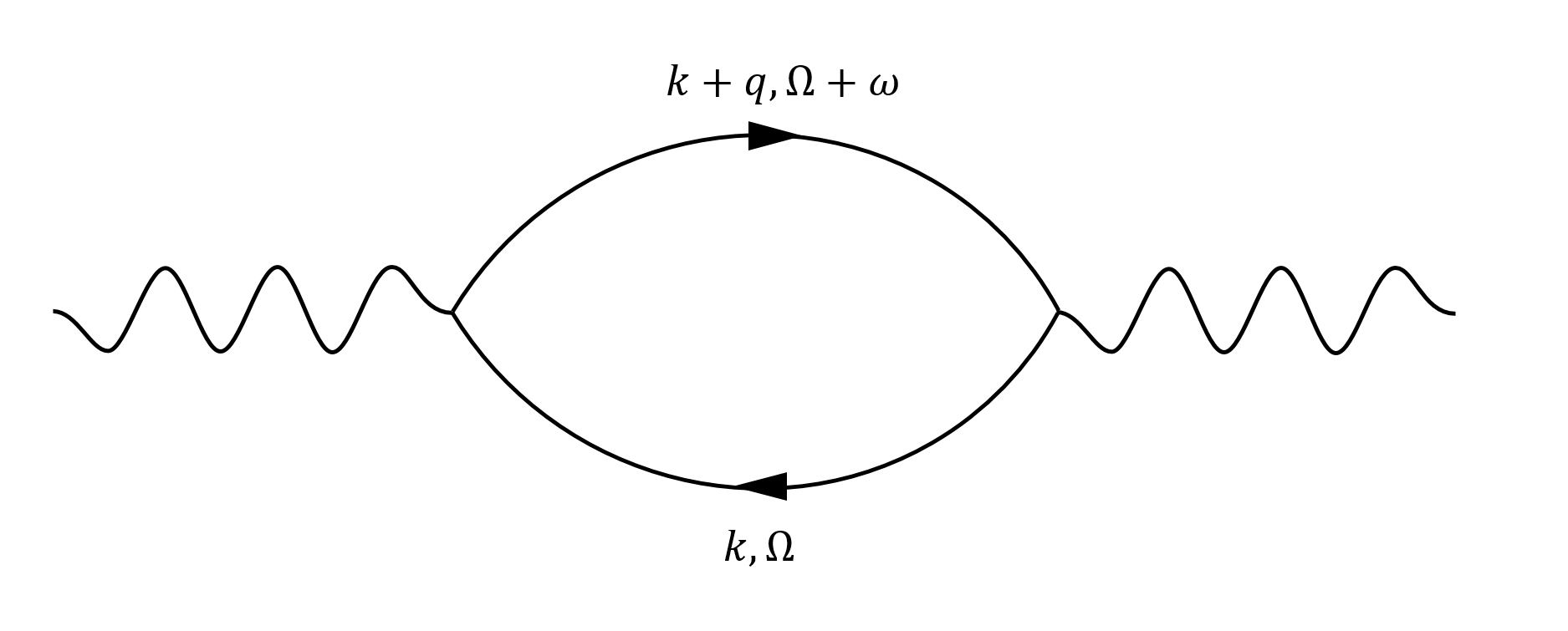}
    \caption{The bubble diagram}
\end{figure}
The $a_x a_y$ term can thus be obtained from the bubble diagram in Fig.~\ref{bubble}. This gives
\be
    \delta{\cal L}^{aa} = \frac{1}{2}\int \frac{d^2\v q d\omega}{(2\pi)^3} \: k^{aa}_{ij}(\v q,\omega) a_i(\v q,\omega)a_j(-\v q,-\omega)
\ee
where
\be
\label{bbl}
    k^{aa}_{ij}(\v q, \omega) = \int \frac{d^2\v k d\Omega}{(2\pi)^3} \: \text{Tr}\bigg[G(\v k,\Omega)j_i(\v k,\v k+\v q)G(\v k +\v q,\Omega+\omega)j_j(\v k+\v q,\v k)\bigg],
\ee
Here the Green's function is
\be
\label{Gkw}
    G(\v k,\Omega) = \sum_n \frac{\ket{u_{n,\v k}}\bra{u_{n,\v k}}}{i\Omega-E_{n\v k}}
\ee
where $n$ is the band index, and  the ``right'' current operator is given by
\be
    j_i(\v k,\v k+\v q) = \sum_\v G \frac{k_i+G_i+q_i/2}{m^*}e^{\frac{i}{2}(\v k+\v G)\times \v q}\ket{\v k+\v G+\v q}\bra{\v k+\v G}
\ee
We are interested in $k^{aa}_{ij}(\omega, \v q=0)\equiv k^{aa}_{ij}(\omega)$.
Then Eqn.~\ref{bbl} becomes
\be
\label{bbl2}
    k^{aa}_{ij}(\omega) = \int \frac{d^2\v k d\Omega}{(2\pi)^3} \: \text{Tr}\bigg[G(\v k,\Omega)j_i(\v k)G(\v k,\Omega+\omega)j_j(\v k)\bigg]
\ee
and the current
\be
\begin{split}
    j_i(\v k) \equiv j_i(\v k,\v k)
    &= \frac{\partial \epsilon^0_{\v k+\v G}}{\partial k_i}\ket{\v k+\v G}\bra{\v k+\v G}= \frac{\partial H_0}{\partial k_i}\\
    &= \sum_\v G \frac{k_i+G_i}{m^*}\ket{\v k+\v G}\bra{\v k+\v G}
\end{split}
\ee

In the following we will work in the Bloch basis. Let $\ket{\psi_{n,\v k}}$ be the eigenstate of the Hamiltonian $H(\v k)$, where $n$ is the band index. (In Appendix.~\ref{Appendix:Chern} and \ref{appendix:numerical}, $n$ was set to 0 representing the lowest band and dropped for simplicity.) Namely, we have
\be
    \ket{\psi_{n, \v k}} = \sum_\v G u_{n,\v G}(\v k)\ket{\v k+\v G}
\ee
where $\ket{\v k+\v G}$ is the plane wave state. 
Denote by $\v r$  the generator of momentum space translation,
\be
    \v r = \int d^2\v k \sum_\v G \ket{\v k+\v G}(i\partial_\v k)\bra{\v k+\v G}
\ee
so that,
\be
    e^{-i\v q\v r} = \int d^2\v k \sum_\v G  \ket{\v k-\v q+\v G}\bra{\v k+\v G}
\ee
The corresponding Bloch function $\ket{u_{n,\v k}}$ is 
\be
    \ket{u_{n\v k}} = e^{-i\v k\v r}\ket{\psi_{n,\v k}} = \sum_\v G u_{n,\v G}(\v k)\ket{\v G}
\ee
Using  Eqn.~\ref{Gkw}, we obtain the Bloch basis Green's function as
\be
\label{Gt}
    \tilde{G}(\v k, \Omega) = e^{-i\v k\v r} G(\v k,\Omega) e^{i\v k\v r} = \sum_n {e^{-i\v k\v r} \frac{\ket{\psi_{n\v k}}\bra{\psi_{n\v k}}}{i\Omega -E_{n\v k}} e^{i\v k\v r} } = \sum_n { \frac{\ket{u_{n\v k}}\bra{u_{n\v k}}}{i\Omega -E_{n\v k}}}
\ee
where $\ket{u_{n\v k}}$ is the Bloch vector that diagonalizes the Hamiltonian in the Bloch basis
\be
\label{Ht}
\begin{split}
    \tilde{H}(\v k) &= e^{-i\v k\v r} H(\v k) e^{i\v k\v r}\\
    &= \sum_{\v G,\v G'} e^{-i\v k\v r} \ket{\v k+\v G} \bra{\v k+\v G}H\ket{\v k+\v G'} \bra{\v k+\v G'}e^{i\v k\v r}\\
    &= \sum_{\v G,\v G'} \ket{\v G} \bra{\v k+\v G}H\ket{\v k+\v G'} \bra{\v G'}
\end{split}
\ee
The current operator in the Bloch basis is
\be
\label{vt}
\begin{split}
    \tilde{j}_i(\v k,\v k+\v q)
    &= e^{-i(\v k+\v q)\v r} v_i(\v k,\v k+\v q) e^{i\v k\v r}\\
    &= \sum_\v G \frac{k_i+G_i+q_i/2}{m^*}e^{\frac{i}{2}(\v k+\v G)\times \v q}\ket{\v G}\bra{\v G}
\end{split}
\ee
Using Eqn.~\ref{Gt} and \ref{vt}, we rewrite the response from the bubble diagram in Eqn.~\ref{bbl} as
\be
\label{bblt}
    k^{aa}_{ij}(\v q, \omega) = \int \frac{d^2\v k d\Omega}{(2\pi)^3} \: \text{Tr}\bigg[\tilde{G}(\v k,\Omega)\tilde{j}_i(\v k,\v k+\v q)\tilde{G}(\v k +\v q,\Omega+\omega)\tilde{j}_j(\v k+\v q,\v k)\bigg]
\ee
Taking $\v q=0$ and expanding to leading order in $\omega$, this becomes
\be
\label{bblt2}
    k^{aa}_{ij}(\omega) = \omega \int \frac{d^2\v k d\Omega}{(2\pi)^3} \: \text{Tr}\bigg[\tilde{G}(\v k,\Omega)\tilde{j}_x(\v k)\partial_\Omega \tilde{G}(\v k,\Omega)\tilde{j}_y(\v k)\bigg]
\ee
where
\be
\label{jt}
    \tilde{j}_i(\v k) \equiv \tilde{j}_i(\v k,\v k) = \sum_\v G \frac{k_i+G_i}{m^*}\ket{\v G}\bra{\v G} =\frac{\partial \tilde{H}_0(\v k)}{\partial k_i}
\ee
and
\be
    \tilde{H}_0(\v k) = e^{-i\v k\v r} H_0(\v k) e^{i\v k\v r} = e^{-i\v k\v r} \sum_{\v G} \epsilon^0_{\v k+\v G} \ket{\v k+\v G} \bra{\v k+\v G} e^{i\v k\v r} = \sum_{\v G} \epsilon^0_{\v k+\v G} \ket{\v G} \bra{\v G}
\ee
Now the ``right'' density operator in the Bloch basis is
\be
\begin{split}
    \tilde{\tilde{\rho}}^R_{\v q}(\v k) = \tilde{\tilde{\rho}}^R(\v k- \v q, \v k)
    &= e^{-i(\v k-\v q)\v r}\tilde{\rho}^R_{\v k-\v q, \v k}e^{i\v k\v r}\\
    &= e^{-i(\v k-\v q)\v r}\sum_\v G \ket{\v k-\v q+\v G}e^{-\frac{i}{2}(\v k+ \v G)\times \v q}\bra{\v k+\v G}e^{i\v k\v r}\\
    &= \sum_\v G \ket{\v G}e^{-\frac{i}{2}(\v k+ \v G)\times \v q}\bra{\v G}
\end{split}
\ee
which is diagonal with the property $\tilde{\tilde{\rho}}^R(\v k', \v k) = \tilde{\tilde{\rho}}^R(\v k, \v k')^\dagger$.
One can show that
\be
\label{HVt}
    \tilde{\tilde{\rho}}^R(\v k-\v q, \v k) \tilde{H}_V(\v k) \tilde{\tilde{\rho}}^R(\v k, \v k-\v q) = \tilde{H}_V(\v k-\v q)
\ee
Next we expand $\tilde{\tilde{\rho}}^R_{\v q}(\v k)$ around $\v q=0$, and introduce the polarization operator $\tilde{P}_i$
\be
\label{rhoexp}
    \tilde{\tilde{\rho}}^R_{\v q}(\v k) = \tilde{\tilde{\rho}}^R(\v k-\v q, \v k) = 1+iq_i\tilde{P}_i(\v k)+O(q^2)
\ee
Therefore, the polarization is
\be
    \tilde{P}_i(\v k) = \sum_{\v G}\frac{1}{2}\epsilon_{ij}(\v k+\v G)_j \ket{\v G}\bra{\v G}
\ee
which is a diagonal and hermitian operator. Using Eqn.~\ref{rhoexp}, we can expand Eqn.~\ref{HVt} as
\be
    \tilde{H}_V(\v k) + iq_i[\tilde{P}_i(\v k),\tilde{H}_V(\v k)] = \tilde{H}_V(\v k) - q_i \frac{\partial \tilde{H}_V(\v k)}{\partial k_i} 
\ee
Thus we have
\be
\label{current-polarization}
    \frac{\partial \tilde{H}_V(\v k)}{\partial k_i}  = -i[\tilde{P}_i(\v k),\tilde{H}_V(\v k)] = -i[\tilde{P}_i(\v k),\tilde{H}(\v k)]
\ee
where in the last identity we utilized the fact that $[P_i(\v k),\tilde{H}_0(\v k)]=0$, since both of them are diagonal in the Bloch basis. Then Eqn.~\ref{jt} is rewritten as
\be
\label{vit}
    \tilde{j}_i(\v k) = \frac{\partial \tilde{H}_0(\v k)}{\partial k_i} = \frac{\partial \tilde{H}(\v k)}{\partial k_i} - \frac{\partial \tilde{H}_V(\v k)}{\partial k_i} =  \frac{\partial \tilde{H}(\v k)}{\partial k_i} + i[\tilde{P}_i(\v k),\tilde{H}(\v k)] =[\partial_i+i\tilde{P}_i(\v k),\tilde{H}(\v k)]
\ee

Finally we are in position to compute the bubble diagram contribution Eqn.~\ref{bblt}. Plugging Eqn.~\ref{Gkw} and \ref{vit} into Eqn.~\ref{bblt} and evaluating the frequency integral, we find
\be
\label{kxyA1}
\begin{split}
    k^{aa}_{ij}(\omega) = i\omega\int \frac{d^2\v k}{(2\pi)^2}&\sum_{n,m} \frac{\theta(-E_{n\v k})-\theta(-E_{m\v k})}{(E_{n\v k}-E_{m\v k})^2}\\
    &\bra{u_{n\v k}}[\tilde{H}(\v k),(\partial_i +i\tilde{P}_i(\v k))]\ket{u_{m\v k}}\bra{u_{m\v k}}[\tilde{H}(\v k),(\partial_j +i\tilde{P}_j(\v k))]\ket{u_{n\v k}}
\end{split}
\ee
Obviously, this object is antisymmetric, namely $k^{aa}_{ij}(\omega)=-k^{aa}_{ji}(\omega)$, which is exactly what we expect. Therefore,
\be
\label{kAaa1}
    k^{aa}_{xy}(\omega) = k^{A,aa}_{xy}(\omega) = \frac{1}{2} \epsilon_{ij}k^{aa}_{ij}(\omega)
\ee
With this we can further simply Eqn.~\ref{kxyA1}. First, for the $\theta(-E_{m\v k})$ term, we replace $m\leftrightarrow n$, interchange $i,j$, and utilize $\tilde{H}(\v k)\ket{u_{n\v k}} = E_{n\v k}\ket{u_{n\v k}}$,
\be
\label{kxyA2}
    k^{aa}_{xy}(\omega) = i\omega\epsilon_{ij}\int \frac{d^2\v k}{(2\pi)^2}\sum_n\sum_{m\neq n} \theta(-E_{n\v k})\bra{u_{n\v k}}(\partial_i +i\tilde{P}_i(\v k))\ket{u_{m\v k}}\bra{u_{m\v k}}(\partial_j +i\tilde{P}_j(\v k))\ket{u_{n\v k}}
\ee
We can extend the summation over $m$ in Eqn.~\ref{kxyA2} to include $m=n$, since the added term is zero under exchange of $i,j$ indices. Therefore, write
\be
\label{kxyAt}
    k^{aa}_{xy}(\omega) = i\omega\epsilon_{ij}\int \frac{d^2\v k}{(2\pi)^2}\sum_{n,m} \theta(-E_{n\v k})\bra{u_{n\v k}}(\partial_i +i\tilde{P}_i(\v k))\ket{u_{m\v k}}\bra{u_{m\v k}}(\partial_j +i\tilde{P}_j(\v k))\ket{u_{n\v k}}
\ee
Eqn.~\ref{kxyAt} expands into four terms. We evaluate them one by one, starting with the derivative term
\be
\label{ddt}
\begin{split}
    &i\omega\epsilon_{ij}\int \frac{d^2\v k}{(2\pi)^2}\sum_{m} \bra{u_{0\v k}}\partial_i\ket{u_{m\v k}}\bra{u_{m\v k}}\partial_j\ket{u_{0\v k}}\\
    =&-i\omega\epsilon_{ij}\int \frac{d^2\v k}{(2\pi)^2}\sum_{m} \braket{\partial_iu_{0\v k}|u_{m\v k}}\braket{u_{m\v k}|\partial_ju_{0\v k}}\\
    =&-i\omega\epsilon_{ij}\int \frac{d^2\v k}{(2\pi)^2} \bra{\partial_iu_{0\v k}}\partial_j\ket{u_{0\v k}}\\
    =&\omega\int \frac{d^2\v k}{(2\pi)^2} \nabla \times A(\v k) =  \frac{\omega C}{2\pi} = -\frac{\omega}{2\pi}\\
\end{split}
\ee
where in the last line we used the Berry curvature $\v A(\v k) =-i\bra{u_k}\nabla\ket{u_k}$. The polarization term is
\be
\label{PPt}
\begin{split}
    &i\omega\epsilon_{ij}\int \frac{d^2\v k}{(2\pi)^2}\sum_{m} \bra{u_{0\v k}}i\tilde{P}_i(\v k)\ket{u_{m\v k}}\bra{u_{m\v k}}i\tilde{P}_j(\v k)\ket{u_{0\v k}}\\
    =&-i\omega\epsilon_{ij}\int \frac{d^2\v k}{(2\pi)^2} \bra{u_{0\v k}}\tilde{P}_i(\v k)\tilde{P}_j(\v k)\ket{u_{0\v k}}=0
\end{split}
\ee
since $\tilde{P}_i(\v k)$ and $\tilde{P}_j(\v k)$ commute.
The two mixed terms
\be
\label{D25}
\begin{split}
    &i\omega\epsilon_{ij}\int \frac{d^2\v k}{(2\pi)^2}\sum_{m} \bra{u_{0\v k}}\partial_i\ket{u_{m\v k}}\bra{u_{m\v k}}i\tilde{P}_j(\v k)\ket{u_{0\v k}}+\bra{u_{0\v k}}i\tilde{P}_i(\v k)\ket{u_{m\v k}}\bra{u_{m\v k}}\partial_j\ket{u_{0\v k}}\\
    =&i\omega\epsilon_{ij}\int \frac{d^2\v k}{(2\pi)^2}\sum_{m} -\braket{\partial_iu_{0\v k}|u_{m\v k}}\bra{u_{m\v k}}i\tilde{P}_j(\v k)\ket{u_{0\v k}}+\bra{u_{0\v k}}i\tilde{P}_i(\v k)\ket{u_{m\v k}}\bra{u_{m\v k}}\partial_j\ket{u_{0\v k}}\\
    =&i\omega\epsilon_{ij}\int \frac{d^2\v k}{(2\pi)^2} -\bra{\partial_iu_{0\v k}}i\tilde{P}_j(\v k)\ket{u_{0\v k}}-\bra{u_{0\v k}}i\tilde{P}_j(\v k)\ket{\partial_iu_{0\v k}}\\
    =&\omega\epsilon_{ij}\int \frac{d^2\v k}{(2\pi)^2} \partial_i\big(\bra{u_{0\v k}}\tilde{P}_j(\v k)\ket{u_{0\v k}}\big)-\bra{u_{0\v k}}\partial_i(\tilde{P}_j(\v k))\ket{u_{0\v k}}\\
\end{split}
\ee
The first term of Eqn.~\ref{D25} is
\be
\label{dpt1}
\begin{split}
    \omega\epsilon_{ij}\int \frac{d^2\v k}{(2\pi)^2}\partial_i\bigg( \bra{u_{0\v k}}\tilde{P}_j(\v k)\ket{u_{0\v k}}\bigg) &= \frac{\omega}{2}\epsilon_{ij}\epsilon_{jl}\int d^2\v k \partial_i\bigg(\sum_{\v G}(\v k+\v G)_l \braket{u_{0\v k}|\v G}\braket{\v G|u_{0\v k}}\bigg)\\
    &= -\frac{\omega}{2}\int \frac{d^2\v k}{(2\pi)^2} \partial_i\bigg(\sum_{\v G}(\v k+\v G)_i |u_{0\v k}(\v G)|^2\bigg)\\
    &= -\frac{\omega}{2}\int \frac{d^2\v k}{(2\pi)^2} \nabla \cdot \v K(\v k) = 0
\end{split}
\ee
In the last step we utilize the fact that $K_j(\v k) = \sum_{\v G} (\v k+\v G)_j |u_{m\v G}(\v k)|^2$ is periodic in the Brillouin zone, so the boundary term vanishes. The second term in Eqn.~\ref{D25} is
\be
\label{dpt2}
\begin{split}
    -\omega\epsilon_{ij}\int \frac{d^2\v k}{(2\pi)^2} \bra{u_{0\v k}}(\partial_i\tilde{P}_j(\v k))\ket{u_{0\v k}}
    &= -\frac{\omega}{2}\epsilon_{ij}\epsilon_{jl}\int d^2\v k \sum_{\v G}(\partial_i(\v k+\v G)_l) \braket{u_{0\v k}|\v G}\braket{\v G|u_{0\v k}}\\
    &= -\frac{\omega}{2}\epsilon_{ij}\epsilon_{ji}\int \frac{d^2\v k}{(2\pi)^2} \sum_{\v G}|u_{0\v k}(\v G)|^2\\
    &= \frac{\omega}{2\pi}
\end{split}
\ee
Combining Eqn.~\ref{ddt}, \ref{PPt}, \ref{dpt1} and \ref{dpt2}, we find that
\be
    \lim_{\omega \to 0}\frac{k^{aa}_{xy}(\omega)}{\omega} = 0
\ee
It follows then that the Chern-Simons level $k=0$ for the internal gauge field 

\subsection{Chern-Simons terms involving external gauge field}
The mean field Hamiltonian is 
\be
    H^{(0)}_{MF} = H_0+H_V = \sum_{\v k,\v G} \epsilon^0_{\v k+\v G} c^\dagger_{\v k+\v G} c_{\v k+\v G} + \sum_{\v k, \v G, \v Q} V_\v Q c^\dagger_{\v k+\v G+\v Q} c_{\v k+\v G}e^{-\frac{i}{2}(\v k+\v G)\times\v Q} 
\ee
where $\epsilon^0_\v k = \frac{\v k^2}{2m^*}$. The external gauge field couples through both terms. Through the kinetic term we have
\be
    H^{(1)}_{0} = \sum_{\v q,\v k,\v G}\frac{k_i+G_i+q_i/2}{m^*}A^i_{\v q}e^{-\frac{i}{2}(\v k+\v G) \times \v q}c^\dagger_{\v k+\v G+\v q}c_{\v k+\v G}
\ee
The coupling through the potential term is
\be
    H^{(0)}_{V} + H^{(1)}_{V} = \int d^2\v x \: \bar{c}(\v x) * V(\v x-\v z \times \v A(\v x)) * c(\v x)
\ee
In momentum space, when expanded to first order in $A$ this becomes
\be
\begin{split}
    H^{(0)}_{V} &= \int d^2\v k \sum_{\v G, \v Q} V_\v Q e^{-\frac{i}{2}(\v k+\v G) \times \v Q} c^\dagger_{\v k+\v G+ \v Q}c_{\v k+\v G}\\
    H^{(1)}_{V} &= i\epsilon_{ij}\int d^2\v k d^2\v q A_{j,\v q} \sum_{\v G, \v Q} Q_iV_\v Q e^{-\frac{i}{2}(\v k+\v G) \times \v Q} c^\dagger_{\v k+ \v q + \v G+ \v Q}c_{\v k+\v G}\\
\end{split}
\ee
The current contributed by this coupling is
\be
\begin{split}
    J^V_{j,\v q} &= \frac{\partial H^{(1)}_{V}}{\partial A_{j,\v q}} = i\epsilon_{ij}\int d^2\v k \sum_{\v G, \v Q} Q_iV_\v Q e^{-\frac{i}{2}(\v k+\v G) \times \v Q} c^\dagger_{\v k+ \v q + \v G+ \v Q}c_{\v k+\v G}\\
\end{split}
\ee
Now we take the limit $\v q = 0$ and transform to Bloch basis. This then becomes (in first quantization)
\be
    \tilde{J}^V_{j}(\v k) = e^{-i\v k \v r}J^V_{j,\v q=0}(\v k)e^{i\v k \v r} = i\epsilon_{ij} \sum_{\v G, \v Q} Q_iV_\v Q e^{-\frac{i}{2}(\v k+\v G) \times \v Q} \ket{\v G+ \v Q}\bra{\v G}\\
\ee
Meanwhile, we have
\be
    \tilde{H}_V(\v k) = e^{-i\v k \v r}H_{V}(\v k)e^{i\v k \v r} = \sum_{\v G, \v Q} V_\v Q e^{-\frac{i}{2}(\v k+\v G) \times \v Q} \ket{\v G+ \v Q}\bra{\v G}\\
\ee
Therefore the current is related to the Hamiltonian through (using Eqn.~\ref{current-polarization})
\be
    \tilde{J}^V_{j}(\v k) = 2\frac{\partial \tilde{H}_V(\v k)}{\partial k_j} = \frac{\partial \tilde{H}_V(\v k)}{\partial k_j}-i[\tilde{P}_i(\v k),\tilde{H}(\v k)]\\
\ee
Combining this with the current contributed by the gauge coupling through kinetic term $\tilde{J}^0_{j}(\v k) = \frac{\partial \tilde{H}_0(\v k)}{\partial k_j}$, we get the total ``left'' current
\be
    \tilde{J}_{j}(\v k) = \tilde{J}^0_{j}(\v k) + \tilde{J}^V_{j}(\v k) = \frac{\partial \tilde{H}(\v k)}{\partial k_j}-i[\tilde{P}_i(\v k),\tilde{H}(\v k)] = [\partial_i-i\tilde{P}_i(\v k),\tilde{H}(\v k)]\\
\ee
Now we are ready to calculate the Chern-Simons term from the same bubble diagram as in Fig.~\ref{bubble} but including both internal and external gauge fields. This gives
\be
\begin{split}
    {\cal S}[A,a] = \frac{1}{2}\int \frac{d^2\v q d\omega}{(2\pi)^3} &\frac{d^2\v k d\Omega}{(2\pi)^3} \: \text{Tr}\bigg[G(\v k,\Omega)\bigg(A_i(\v q, \omega)J_i(\v k,\v k +\v q)+a_i(\v q, \omega)j_i(\v k, \v k +\v q)\bigg)\\
    &G(\v k +\v q,\Omega+\omega)\bigg(A_j(-\v q, -\omega)J_j(\v k +\v q, \v k)+a_j(-\v q, -\omega)j_j(\v k +\v q, \v k)\bigg)\bigg]\\
    = \frac{1}{2}\int \frac{d^2\v q d\omega}{(2\pi)^3} &\:
     k^{Aa}_{ij}(\v q,\omega) A_i(\v q,\omega)a_j(-\v q,-\omega)\\
     + & k^{AA}_{ij}(\v q,\omega) A_i(\v q,\omega)A_j(-\v q,-\omega) + k^{aa}_{ij}(\v q,\omega)a_i(\v q,\omega)a_j(-\v q,-\omega)
\end{split}
\ee
We begin by considering the cross-term involving both $A$ and $a$. The contribution at $\v q =0$ to this term is
\be
\begin{split}
    k^{Aa}_{ij}(\omega) A_i(\omega)a_j(-\omega)
    =& \int \frac{d^2\v k d\Omega}{(2\pi)^3} \:
    \text{Tr}\bigg[\tilde{G}(\v k,\Omega)\tilde{J}_i(\v k)\tilde{G}(\v k,\Omega+\omega)\tilde{j}_j(\v k)\bigg]A_i(\omega)a_j(-\omega)\\
    &\qquad\qquad+\text{Tr}\bigg[\tilde{G}(\v k,\Omega)\tilde{j}_j(\v k)\tilde{G}(\v k,\Omega-\omega)\tilde{J}_i(\v k)\bigg]a_j(-\omega)A_i(\omega)\\
    =& \int \frac{d^2\v k d\Omega}{(2\pi)^3} \: \text{Tr}\bigg[\tilde{G}(\v k,\Omega)\tilde{J}_i(\v k)\partial_\Omega \tilde{G}(\v k,\Omega)\tilde{j}_j(\v k)\\
    &\qquad\qquad-\tilde{G}(\v k,\Omega)\tilde{j}_j(\v k) \partial_\Omega \tilde{G}(\v k,\Omega)\tilde{J}_i(\v k)\bigg] \omega A_i(\omega)a_j(-\omega)\\
    =& 2\int \frac{d^2\v k d\Omega}{(2\pi)^3} \: \text{Tr}\bigg[\tilde{G}(\v k,\Omega)\tilde{J}_i(\v k)\partial_\Omega \tilde{G}(\v k,\Omega)\tilde{j}_j(\v k)\bigg] \omega A_i(\omega)a_j(-\omega)\\
\end{split}
\ee
where we have expanded $k^{Aa}_{ij}(\omega)$ to linear order in $\omega$.
Carrying out the frequency integral, we find
\be
\label{kAa}
\begin{split}
    k^{Aa}_{ij}(\omega) &= 2i\omega\int \frac{d^2\v k}{(2\pi)^2}\sum_{n\neq m} \frac{\theta(-E_{n\v k})-\theta(-E_{m\v k})}{(E_{n\v k}-E_{m\v k})^2}\bra{u_{n\v k}}\tilde{J}_i(\v k)\ket{u_{m\v k}}\bra{u_{m\v k}}\tilde{j}_j(\v k)\ket{u_{n\v k}}\\
    & =2i\omega\int \frac{d^2\v k}{(2\pi)^2}\sum_{n\neq m} \frac{\theta(-E_{n\v k})}{(E_{n\v k}-E_{m\v k})^2}\\
    &\qquad\qquad\qquad\bigg(\bra{u_{n\v k}}\tilde{J}_i(\v k)\ket{u_{m\v k}}\bra{u_{m\v k}}\tilde{j}_j(\v k)\ket{u_{n\v k}} -\bra{u_{n\v k}}\tilde{j}_j(\v k)\ket{u_{m\v k}}\bra{u_{m\v k}}\tilde{J}_i(\v k)\ket{u_{n\v k}} \bigg)\\
    & =2i\omega\int \frac{d^2\v k}{(2\pi)^2}\sum_{m} \bigg(\bra{u_{0\v k}}(\partial_i -i\tilde{P}_i(\v k))\ket{u_{m\v k}}\bra{u_{m\v k}}(\partial_j +i\tilde{P}_j(\v k))\ket{u_{0\v k}}\\
    &\qquad\qquad\qquad\qquad\qquad\qquad-\bra{u_{0\v k}}(\partial_j +i\tilde{P}_j(\v k))\ket{u_{m\v k}}\bra{u_{m\v k}}(\partial_i -i\tilde{P}_i(\v k))\ket{u_{0\v k}}\bigg)
\end{split}
\ee
Again we expand this into three parts and simplify them one by one. The pure derivative term in Eqn.~\ref{kAa} is
\be
\begin{split}
    &2i\omega\int \frac{d^2\v k}{(2\pi)^2}  \bigg(-\braket{\partial_iu_{0\v k}|\partial_ju_{0\v k}}
    +\braket{\partial_ju_{0\v k}|\partial_iu_{0\v k}}\bigg)\\
    =& 2\omega\int \frac{d^2\v k}{(2\pi)^2}  \bigg(\partial_iA_j(\v k)-\partial_jA_i(\v k)\bigg) = -2\epsilon_{ij}\frac{\omega}{2\pi} 
\end{split}
\ee
The polarization term vanishes since $\[P_i(\v k),P_j(\v k)\]=0$. The other terms get simplified to
\be
\begin{split}
    &2\omega\int \frac{d^2\v k}{(2\pi)^2}  \sum_m \bra{u_{0\v k}}\tilde{P}_i(\v k)\ket{u_{m\v k}} \bra{u_{m\v k}}\partial_j\ket{u_{0\v k}} - \bra{u_{0\v k}}\partial_i\ket{u_{m\v k}} \bra{u_{m\v k}}\tilde{P}_j(\v k)\ket{u_{0\v k}}\\
    & \qquad\qquad\qquad +\bra{u_{0\v k}}\tilde{P}_j(\v k)\ket{u_{m\v k}} \bra{u_{m\v k}}\partial_i\ket{u_{0\v k}} - \bra{u_{0\v k}}\partial_j\ket{u_{m\v k}} \bra{u_{m\v k}}\tilde{P}_i(\v k)\ket{u_{0\v k}}\\
    =&2\omega\int \frac{d^2\v k}{(2\pi)^2}  \bra{u_{0\v k}}\tilde{P}_i(\v k)\partial_j\ket{u_{0\v k}} + \bra{\partial_iu_{0\v k}}\tilde{P}_j(\v k)\ket{u_{0\v k}}\\
    & \qquad\qquad +\bra{u_{0\v k}}\tilde{P}_j(\v k)\partial_i\ket{u_{0\v k}} + \bra{\partial_ju_{0\v k}}\tilde{P}_i(\v k)\ket{u_{0\v k}}\\
    =&2\omega\int \frac{d^2\v k}{(2\pi)^2}  \partial_j(\bra{u_{0\v k}}\tilde{P}_i(\v k)\ket{u_{0\v k}}) - \bra{u_{0\v k}}(\partial_j\tilde{P}_i(\v k))\ket{u_{0\v k}}\\
    &\qquad\qquad +\partial_i(\bra{u_{0\v k}}\tilde{P}_j(\v k)\ket{u_{0\v k}}) - \bra{u_{0\v k}}(\partial_i\tilde{P}_j(\v k))\ket{u_{0\v k}}\\
    =&-2\omega\int \frac{d^2\v k}{(2\pi)^2} \bra{u_{0\v k}}(\partial_j\tilde{P}_i(\v k)+\partial_i\tilde{P}_j(\v k))\ket{u_{0\v k}}\\
    =&0
\end{split}
\ee
Therefore, we find a non-vanishing coefficient for the mutual Chern-Simons term
\be
    {\cal L}^{Aa} = -\frac{1}{2\pi} \epsilon_{\mu\nu\rho} A_\mu \partial_\nu a_\rho
\ee

In the long wavelength limit, and to quadratic order in $a,A$, this term is gauge invariant under the usual (commutative) gauge transformations of $a,A$. Strictly speaking, as the model has full non-commutative gauge invariance under both `left' and 'right' gauge transformations we should require that this mutual Chern-Simons term be completed by a more elaborate expression, involving higher derivatives and powers of the gauge fields,  that correctly captures these invariances. We have not been able to determine the structure of such an expression. For our purposes of characterizing the long wavelength response of the system to the external gauge field, it is however sufficient to restrict to thsi leading order form. 

In a very similar fashion we obtain the self Chern-Simons term for external gauge field. The contribution from the bubble diagram is
\be
\begin{split}
    k^{AA}_{xy}(\omega) &= \frac{i\omega\epsilon_{ij}}{2}\int \frac{d^2\v k}{(2\pi)^2}\sum_{n\neq m} \frac{\theta(-E_{n\v k})-\theta(-E_{m\v k})}{(E_{n\v k}-E_{m\v k})^2}\bra{u_{n\v k}}\tilde{J}_i(\v k)\ket{u_{m\v k}}\bra{u_{m\v k}}\tilde{J}_j(\v k)\ket{u_{n\v k}}\\
    & =i\omega\epsilon_{ij}\int \frac{d^2\v k}{(2\pi)^2}\sum_{n\neq m} \frac{\theta(-E_{n\v k})}{(E_{n\v k}-E_{m\v k})^2}\bra{u_{n\v k}}\tilde{J}_i(\v k)\ket{u_{m\v k}}\bra{u_{m\v k}}\tilde{J}_j(\v k)\ket{u_{n\v k}} \\
    & =i\omega\epsilon_{ij}\int \frac{d^2\v k}{(2\pi)^2}\sum_{m} \bra{u_{0\v k}}(\partial_i -i\tilde{P}_i(\v k))\ket{u_{m\v k}}\bra{u_{m\v k}}\partial_j -i\tilde{P}_j(\v k))\ket{u_{0\v k}}
\end{split}
\ee
Note that this differs from Eqn.~\ref{kxyAt} only by a negative sign for the mixed terms $\sim \braket{\partial_i}\braket{\tilde{P}_j}$. Therefore in this case both the pure derivative term and the mixed term contribute $-1$ to the Chern-Simons level. Consequently, path integral contribution to the self-Chern-Simons term for external gauge field is
\be
    {\cal L}^{AA} = -\frac{1}{2\pi} \epsilon_{\mu\nu\rho} A_\mu \partial_\nu A_\rho
\ee

\bibliographystyle{apsrev4-2}
\bibliography{LLL}

\end{document}